\documentclass[12pt,preprint]{aastex} 

\usepackage{graphicx}
\usepackage{dcolumn}
\usepackage{bm}

\usepackage{epsfig} 

\newcommand{\cang}{CANGAROO}
\newcommand{\gam}{$\gamma$}
\def\3EG{{3EG J1746-2851}}
\def\p0{{$\pi^0$}}
\def\1018{{$10^{18}$}}

\newcommand{\be}{\begin{equation}}
\newcommand{\ee}{\end{equation}}
\newcommand{\bea}{\begin{eqnarray}}
\newcommand{\eea}{\end{eqnarray}}
\newcommand{\bmu}{\begin{multline}}
\newcommand{\emu}{\end{multline}}
\newcommand{\nn}{\nonumber}

\newcommand{\sae}{Sgr A East\,}

\def\msun{{\,M_\odot}}
\def\lsun{{\,L_\odot}}
\def\simlt{\lower.5ex\hbox{$\; \buildrel < \over \sim \;$}}
\def\simgt{\lower.5ex\hbox{$\; \buildrel > \over \sim \;$}}

\def\gcm3{{\rm\,g\,cm^{-3}}}
\def\ncm3{{\rm\,cm^{-3}}}

\def\>{$>$}
\def\<{$<$}

\lefthead{Crocker et~al.}                      
\righthead{Galactic Center Neutrons}

\received{}
\begin{document}


\title{\bf The AGASA/SUGAR Anisotropies and TeV Gamma Rays from the Galactic
Center: A Possible Signature of Extremely High-energy Neutrons}

\author{Roland M. Crocker$^{1,2}$, Marco Fatuzzo$^{3}$, Randy Jokipii$^{4}$,
Fulvio Melia$^{5}$ 
\\and Raymond R. Volkas$^2$}
\affil{$^1$ Harvard-Smithsonian Center for Astrophysics\\
60 Garden St.,
Cambridge MA 02138\\
rcrocker@cfa.harvard.edu}
\affil{$^2$Research Centre for High Energy Physics, 
School of Physics,
\\The University of Melbourne,
3010 Australia\\
r.crocker, r.volkas@physics.unimelb.edu.au}
\affil{$^{3}$Physics Department,\\Xavier University, Cincinnati, OH 45207\\
fatuzzo@cerebro.cs.xu.edu}
\affil{$^{4}$Department of Planetary Sciences, 
\\The University of Arizona, Tucson, AZ 85721\\
  jokipii@lpl.arizona.edu}
\affil{$^{5}$Physics Department and Steward Observatory, 
\\The University of Arizona, Tucson, AZ 85721\\
melia@physics.arizona.edu}

\begin{abstract}
Recent analysis of data sets from two extensive air shower cosmic ray
detectors shows tantalizing evidence of an anisotropic overabundance of
cosmic rays towards the Galactic Center (GC) that ``turns on'' around
$10^{18}$ eV. We demonstrate that the anisotropy could be due to neutrons
created at the Galactic Center through charge-exchange in proton-proton
collisions, where the incident, high energy protons obey an $\sim E^{-2}$ power
law associated with acceleration at a strong shock. We show that the   
normalization supplied by the gamma-ray signal from EGRET GC source 3EG
J1746-2851 -- ascribed to pp induced neutral pion decay at GeV energies
-- together with a very reasonable spectral index of 2.2, predicts a
neutron flux at $\sim 10^{18}$ eV fully consistent with the extremely high energy
cosmic ray data. Likewise, the normalization supplied by the very recent
GC data from the HESS air-Cerenkov telescope at ~TeV energies is almost
equally-well compatible with the $\sim 10^{18}$ eV cosmic ray data. Interestingly,
however, the EGRET and HESS data appear to be themselves incompatible. We
consider the implications of this discrepancy. We discuss why the Galactic
Center environment can allow diffusive shock acceleration at strong shocks
up to energies approaching the ankle in the cosmic ray spectrum. Finally,
we argue that the shock acceleration may be occuring in the shell of Sagittarius A East,
an unusual supernova remnant located very close to the Galactic Center. If
this connection between the anisotropy and Sagittarius A East could be    
firmly established it would be the first direct evidence for a particular 
Galactic source of cosmic rays up to energies near the ankle.
\end{abstract}



\section{Introduction}

The origin of cosmic rays (CRs) is one of the most important unsolved problems
in astrophysics. While it has long been speculated that diffusive shock 
acceleration of protons and ions at shock fronts associated with supernova 
remnants (SNRs) is the  mechanism likely responsible for accelerating the bulk of
high energy
cosmic rays, definitive
observational proof has been elusive. Further, the conditions at almost all
known SNRs seem not to  promote the acceleration of CRs beyond the `knee' feature 
in the spectrum at $\simeq 5 \times 10^{15}$ eV. The acceleration mechanism for
CRs between the knee and the `ankle' at few $\times 10^{19}$ eV 
therefore seems to be an even deeper puzzle.

The purpose of this paper is to argue that the Galactic Center (GC), a region
with relatively extreme conditions compared to the rest of the Milky Way,
is a likely site where CRs are accelerated up to the ankle. Our argument
is based on the following: 
(i) The EGRET $\gamma$-ray source 3EG J1746-2851,
most likely located near the GC, provides good evidence for pion production from high
energy proton-proton (pp) collisions. 
Neutrons will then inevitably also be
produced by this source. 
(ii) The TeV \gam -rays from the direction of the GC
detected by a number of air Cerenkov telescopes, in particular, the HESS instrument,
also support the notion that high-energy proton acceleration and collision
processes (again, leading inevitably to neutron production) are occuring in this region.
(iii) The AGASA CR anisotropy for the
energy range $10^{17.9} - 10^{18.5}$ eV is consistent with high energy neutron
emission from the GC. 
(iv) The reanalysed SUGAR data also reveal an anisotropy
for this energy regime from a direction close to the GC.
(v) The SNR \sae provides a plausible specific
GC system where hadronic acceleration up to the ankle can occur, differing as it does
from other Galactic SNRs by the special conditions of its GC environment.

If the connection between the anisotropies and \sae could be firmly established,
it would be the first identification of a specific source producing high energy, Galactic CRs,
and, moreover, it would be proven to be an important, possibly unique, contributor to
CR acceleration up to the ankle. The southern-hemisphere AUGER detector, currently
under construction, will test our hypothesis in the relatively near future: it should
see a significant point source of $\sim 10^{18}$ eV neutrons at the GC. (For lower
energies, previous work has shown that a GC neutrino signal should also be seen
by a northern-hemisphere km$^3$ neutrino telescope: 
Crocker et~al. 2002).

We extend earlier work on GC CR production in three important ways. First, postulating
a GC source of high energy protons obeying an $\sim E^{-2}$ power law at the source, 
with normalisation
fixed by the concomitant (GeV) EGRET $\gamma$-ray observations of 3EG J1746-2851, we
calculate neutron production through charge-exchange $pp \to nX$ reactions. {\it Quite
non-trivially, we find neutron fluxes consistent with the magnitudes of the AGASA
and SUGAR anisotropies.} Second, we likewise show that a simultaneous fit to the (TeV)
data supplied by the HESS instrument and the cosmic ray anisotropy data
is also consistent with the existence of a population of of shock-accelerated protons
governed by a $\sim E^{-2}$ power law up to extremely high energies.
Third, we show that diffusive shock acceleration beyond the knee can occur
provided that there is a significant magnetic field component {\it perpendicular} to the SNR shock
propagation direction (i.e., parallel to the shock front). 
We explain why the special conditions at the GC, especially the higher density
of the ambient medium and higher magnetic field, 
can realise this situation. Third, we argue that the specific
GC SNR \sae is the most likely specific candidate site for CR acceleration up to the ankle.
Note, however, that it is not necessary to identify the specific source for
the conclusion italicised above to follow.

Note also that we admit from the start that our model is not 
consistent with all available data. The data, however, are 
inconsistent amongst themselves in the two important instances where
there is disagreement with our model (see \S \ref{sectn_noncoin} and \S \ref{section_TeV}). 
These instances of disagreement -- both discussed in further 
detail below -- are (i) that the SUGAR results indicate a point source
offset by $7.5^\circ$ from the GC (whereas we predict, of course, 
a source at the
position of the GC on the sky) and (ii) that the $\sim$ TeV measurements 
of the \gam -ray flux from the GC (by a number of instruments)
appears to be deficient compared 
to what we would expect. We discuss a number of possible resolutions of this discrepancy,
finding the most favorable to hinge on there being two effective GC sources of \gam -rays,
an idea which has some support from current observations.

\section{Origin of Cosmic Rays: Role of Neutrons}

The hypothesis that shocks at SNRs
are responsible for the acceleration of CRs
over the bulk of the observed spectrum
is fifty years old \citep{Shklovskii1953}.
And, indeed,
there is strong, albeit circumstantial, evidence
that SNRs do accelerate cosmic rays
up to, at least, the `knee' in the spectrum at 
a few PeV (we set $E_\mathrm{knee} \equiv 10^{15}$ eV for definiteness).
This evidence comes primarily from two arguments, viz,
\begin{enumerate}
\item Supernovae seem to be the only
class of Galactic object able to inject the power necessary to maintain 
the observed cosmic ray output of about $10^{48}$ 
ergs/year 
(e.g., Longair (1994)).
\item The cosmic ray spectrum is
governed by a power law with spectral index around
2.7. This is close to the universal power law
of spectral index $\sim$ 2 theoretically expected
from diffusive shock acceleration at the sort of strong shock
associated with a supernova blast wave.
(Note that this theoretical expectation for spectral indices close
to two -- at the source -- 
is observationally confirmed by radio and \gam -ray data from various
SNRs.)
The difference between these two indices, further, can
be compellingly explained
as arising from energy-dependent propagation/confinement
effects. 
\end{enumerate}

However, there is yet no {\it direct} observational 
evidence for the SNR-CR connection and certainly no
particular SNR has been proven to be
a CR source.
Further, many researchers have found that their models are
pushed to accelerate particles up to $E_\mathrm{knee}$.
This becomes doubly troubling given the fact that, as emphasized
by Jokipii and Morfill (1991), matching the spectra at the 
knee requires, short of a cosmic conspiracy, 
that the population of cosmic rays above the
knee is closely related to that below the knee. 
In fact, there are good reasons to think that the bulk
of the cosmic rays are Galactic in origin up to the `ankle' in the
spectrum at a few $\times\ 10$ EeV (1 EeV $\equiv$ $10^{18}$ eV), 
not the least of which is that a proton of 
this energy has a gyroradius the size of the radius of the Galactic
disk.\footnote{Yet higher energy CRs may therefore no longer be confined
to our Galaxy, so are almost certainly extra-Galactic. The up-turn
in the spectrum at the ankle is nicely consistent with a
new population taking over from a diminishing Galactic component.}

We would like, therefore, to determine
whether it is, indeed, the case that SNRs 
can accelerate particles up to 
energies of $10^{18 - 19}$ eV. Further, it would be
helpful to find evidence
that some {\it particular} object is accelerating particles to these
extremely high energies (EHEs).

We obviously need to look for a signal in electrically neutral particles, so that the
location of the source is not scrambled by deflection due to the Galactic magnetic
field.\footnote{Charged particles can still be used for very close sources. See, for
example, 
Chilingarian (2003)
for a study of putative CR emission from the 300 pc
distant Monogem ring SNR.} There are three main candidates: photons, neutrinos and neutrons.
Photons with energy $10^{17,18}$ eV will be produced by the decay of neutral pions produced by
hadronic collisions. However, such EHE \gam 's will interact
strongly with background media long before reaching the Earth, so they are not
useful as direct probes at those energies. The flux of $10^{17,18}$ eV neutrinos,
another concomitant of hadronic collisions, is
expected to be orders of magnitude too small to be seen by future km$^3$ neutrino
telescopes. Although three of us have previously shown that the GC should emit
neutrinos (Crocker et al. 2000; Crocker et al. 2002)
[see also \cite{Alvarez2002}], 
their detection could not settle the question of where the $10^{18}$ eV CRs
are coming from, because the detectable flux of neutrinos lies at considerably 
lower energies.
 



This leaves us with neutrons.
As for neutrinos and \gam -rays, neutrons
are an inescapable concomitant of hadronic acceleration of protons and
ions: charge exchange occurs in a non-negligible
fraction of all interactions
between accelerated protons and ambient protons. Neutrons are also produced
in p\gam \ collisions and dissociations of accelerated ions (see below).
The neutron, however, is unstable in free space, a fact we have to take into
account when contemplating neutron astronomy.


The neutron is the longest-lived unstable `elementary' particle with a 
decay time at rest,
$\tau_n$, of  886 seconds \citep{Hagiwara2002}.
This means that a neutron will travel, 
on average, a distance of
\be
d_n(E_n) = c \gamma_n \tau_n 
\simeq 9 \left(\frac{E_n}{\mathrm{EeV}}\right)\mathrm{kpc},
\label{eqn_mfp}
\ee
(where the Lorentz factor is given by $\gamma_n \equiv E_n/m_n$)
in free space before decaying.
What plausibly neutron-producing regions lie within the $\sim 9$ kpc
distance travelled on average by an EeV neutron?
The Galactic Center -- one of the most energetic regions in the Galaxy -- 
at a distance of around 8.5 kpc is the principal candidate.


What would be a smoking gun signature for EHE CR neutrons?
It is in extensive airshowers (EAS) apparently initiated by particles 
coming from the direction of the GC that one would need to look for GC 
neutrons. The signal might be very difficult
to disentangle, essentially because any neutron-initiated EAS at these
energies is likely to be
indistinguishable from a background, proton-initiated EAS\footnote{Though note here
that there is a theoretical
possibility, at least, that empirical data may one day be able to directly
distinguish EHE cosmic ray neutrons from protons via
characteristic differences between the
$\mu^+$ to $\mu^-$ ratios seen in the extensive air showers
generated by these particles (a proton will produce an
excess of $\pi^+$ and, therefore, $\mu^+$  in the forward region
 whereas a neutron will produce an excess of  $\pi^-$ and, therefore, $\mu^-$).}
 (which proton,
given the presence of the Galactic magnetic field,
might originate from a source considerably away from the GC).
The best evidence
for GC neutrons one might hope for, then,
is an {\it anisotropy} in the EeV cosmic ray data in the form 
of an excess towards the GC.  Intriguingly, there is tantalizing evidence, 
which we now briefly review, from two different data sets that just such
an anisotropy exists.

\section{The Galactic Center Anisotropy Explained by a GC Source of
Protons/Neutrons}

Recent analysis of data from two different cosmic ray detectors
has revealed the presence of an anisotropic overabundance
of cosmic rays coming from the general  direction of the Galactic Center.
Consistent with these findings, analysis of data from a third array
has found a broad anisotropy along the Galactic plane.
We now briefly review each of these findings.

Statistically, the most robust determination for an anisotropy
has been by the Akeno Giant Shower Array (AGASA) Group \citep{Hayashida1999}
which, in analysis of 114,000 airshowers found a strong -- 4\% amplitude --
anisotropy in the energy range $10^{17.9} - 10^{18.3}$ eV (we label
the $10^{17.9}$ eV energy at which the anisotropy apparently `turns on'
$E_\mathrm{onset}$). Note that
the AGASA Collaboration \citep{Takeda1999b} has estimated that the 
systematic uncertainty in its instrument's energy callibration is
30\% and we shall adopt this figure in our analysis.
The group's two-dimensional analysis of the data
showed that this anisotropy could be interpreted as an excess of air showers
from two regions each of $\sim 20^\circ$ extent, one of 4$\sigma$
significance near the GC 
and another of 3$\sigma$ in Cygnus.
Subsequent re-analysis by the AGASA group -- incorporating new data --
has only served to bolster the claim that the anisotropy is real \cite{Hayashida1999b} with,
this time, a 4.5 $\sigma$ excess seen near the GC over a beam size of 20$^\circ$ between
$10^{18.0} - 10^{18.4}$ eV. 

Prompted by the AGASA result, Bellido et al. (2001)
re-analyzed the data collected by the SUGAR cosmic ray detector
which operated from 1968 to 1979 near Sydney. Setting {\it a priori}
an energy range similar to that determined for the AGASA
anisotropy, these researchers also found an anisotropy, consistent
with a point source located at 7.5$^\circ$ from the GC -- and 6$^\circ$
degrees from the AGASA maximum over an energy range of $10^{17.9} - 10^{18.5}$ eV.

Lastly, the HiRes Collaboration has
seen a Galactic Plane enhancement in cosmic ray events
in the energy range between $10^{17.3}$ and $10^{18.5}$ eV
 with 3.2$\sigma$
confidence \citep{Bird1999} (this is consistent with the AGASA and SUGAR
results because the HiRes study was broad scale and did not attempt to pin down
whether any particular Galactic longitudes were responsible for the detected excess:
Bellido et al. 2001).

As we shortly set out, the anisotropies mentioned above have 
what we believe is a natural
explanation in terms of neutron emission from the GC region
(see the subsection on `GC Neutron Models' below).
Before we discuss this idea in detail, however,
we also consider the possibility that the observed anisotropies
can be explained directly by a diffusive flux of charged particles
from the GC. We label such scenarios `GC Charged Particle Models'.

\subsection{GC Charged Particle Models}

There has been a concerted effort to model
the propagation of charged particles
from an assumed source in the vicinity of the
GC, through various assumed configurations of
the Galactic magnetic field, to Earth to see whether
these models can reproduce the observed anisotropies
(Clay 2000;Bednarek, Giller, and Zielinska 2002;Candia, Mollerach, and Roulet 2002).
In general, researchers have found that a fair
degree of correspondence between models and reality can be achieved, with,
in particular, an anisotropy becoming evident 
for ${\cal O}$[EeV] protons propagating through various field configurations
with a magnetic field amplitude of ${\cal O}$[$\mu$G].
Finding an exact correspondence for `turn-on' and `turn-off' energies seems, however,
to require quite some fine-tuning of the particulars of 
${\mathbf B}_\mathrm{Galactic}$ and is not such a {\it robust}
solution as the neutron idea. Further,
for reasonable amplitudes of ${\mathbf B}_\mathrm{Galactic}$ 
modeling studies show that
the actual separation seen between the GC and
the observed excess will exceed the
${\cal O}$[10$^\circ$] observed 
(Candia, Epele, and Roulet 2002).
In fact, Medina-Tanco and Watson (2001) report that the source must be within
$\sim 2$ kpc of the Earth to reproduce the observed deflection.
This is of course, a possibility, but then the near alignment with
the anisotropies and the GC becomes, essentially, a coincidence.
Lastly, and perhaps most tellingly,
the (consistent with-) point-like anisotropy seen in the
SUGAR data is very difficult to reproduce in modeling of charged particle
trajectories  because the particles
tend to smear out until rather higher energies
(for reasonable magnetic fields) than $E_\mathrm{onset}$.

\subsection{GC Neutron Models}

The broad idea that neutron emission from the GC may produce an anisotropy
was, to our knowledge, first mooted by 
Jones (1990).
The neutron idea was 
subsequently revived by the AGASA group in the paper announcing their discovery of the
GC anisotropy \cite{Hayashida1999}. 
The AGASA paper authors pointed out that the anisotropy
`turn on' at a definite energy of $\sim$ EeV finds a natural
explanation in the fact that -- as outlined above --
this energy corresponds to a gamma
factor for neutrons large enough that they can reach us from
the GC. So, broadly, neutrons below this energy
decay in propagation and are then diverted by Galactic magnetic fields.
The cessation of the anisotropy above $\sim 10^{18.4}$ eV can be explained
as either due to a very steep GC source spectrum or an actual cut-off
in the source so that the background takes over again at this energy.

Most recently, the broad idea above has been
considerably refined in the work of Bossa et al. (2003), whose
basic scenario we follow and now explain.
These authors have made detailed propagation calculations
following the trajectories of protons from neutrons that decay in flight
from the GC or, to be precise, they follow the trajectories of
anti-protons leaving the Earth and calculate the probability that
a decay should occur over the interval during which the
anti-proton's path points back to the assumed GC source.
Using this procedure they calculate detailed maps of the 
arrival direction of the combined neutron and proton signal
for various values of an assumed  Galactic magnetic field 
which has both a regular and a turbulent component.
The Bossa et al. (2003) scenario extends that presented by Medina-Tanco
and Watson (2001) in which,
essentially, decay protons produced close to the Earth arrive from directions close to the
GC whereas those produced in the inner Galaxy arrive preferentially from the directions of
the spiral arms -- thus also neatly explaining the Cygnus anisotropy --
as their trajectories wind around the regular magnetic field lines 
(Bossa et al. 2003).

Crucially, the GC, with declination $\delta = -28.9^\circ$, is outside the 
field of view of AGASA (which is limited to $\delta > -24^\circ$; Bossa et al. 2003).
This provides a natural explanation for the `turn off' of the
AGASA anisotropy without the need for a source cut-off: at energies $\gtrsim 10^{18.4}$ eV, neutrons
do not (on average) decay in flight from the GC, but, instead, travel in a straight line
from the GC to Earth to produce a point-like anisotropy {\it out} of the field
of AGASA. Moreover, Bossa et al. (2003) were also able to reproduce
the sharp onset of the
AGASA anisotropy at $10^{17.9}$ eV by positing a Galactic magnetic
field random component of fairly large amplitude
(3 $\mu$G) and also assuming 
a GC source governed by a spectral index of 2.2.
As mentioned above, 
these researchers also relate the Cygnus region excess seen by AGASA
to the GC source with a Galactic magnetic field whose
regular component is along the spiral arms (and not just azimuthal).
Finally, in the Bossa et al. (2003) scenario, the {\it magnitudes
and morphologies} of the 
AGASA and SUGAR results
were shown to be compatible
({\it directional} consistency being problematic -- see below).
In addition to approximate consistency in magnitude,
the fact that the AGASA and SUGAR anisotropies are, respectively,
diffuse and (consistent with) point-like finds a natural 
explanation.
Indeed, from the requirement
that the source normalization generate
the observed 4\% amplitude anisotropy of the right-ascension first
harmonic (and continuing to assume a power-law source
spectrum with index 2.2), Bossa et al. (2003) determine that the total
luminosity of the GC source over the specified decade in energy be
\be
L_{GC}(10^{17.5}-10^{18.5}) \simeq 4 \times 10^{36} \mathrm{erg}\;{\mathrm{s}}^{-1}\;,
\ee
which implies a direct neutron flux between $10^{17.9}-10^{18.5}$ eV
of $2 \times 10^{-17}$ cm$^{-2}$ s$^{-1}$ (no error range given), 
which is not very different from the
SUGAR result of $(9 \pm 3) \times 10^{-18}$ cm$^{-2}$ s$^{-1}$.
Bossa et al. (2003) also point out that these two figures need not
be exactly equal: protons will be, in general, delayed by many thousands of
years with respect to the neutron arrival times so the source intensity need
not be exactly the same when the neutrons and protons we observe today were
separately emitted.

A couple of important points one should note
about the Bossa et al. (2003) scenario (as these researchers themselves 
remark) are that (i) for the magnetic field adopted, the 
phase ($\sim 330^\circ$) of the first harmonic in right ascension found
is somewhat larger than that detected by AGASA and
(ii) because Bossa et al. (2003) {\it do} relate the Cygnus excess to a GC source,
this excess should disappear for $E \gtrsim$ 2 EeV
{\it independently of the intrinsic source energy cut-off} because
any reasonable Galactic magnetic field could not, reasonably, shepherd cosmic rays
above this energy along a spiral arm.

\subsection{Non-coincidence of the SUGAR ansisotropy with the GC}
\label{sectn_noncoin}

That the AGASA ansisotropy is not exactly coincident with the GC
finds a reasonable explanation in the fact already presented that the actual GC is out
of the field of view of this instrument.
That the SUGAR ansisotropy, however, is not coincident with the GC
presents a challenge to all  scenarios that would posit that the source of the EHE
cosmic rays is at the GC.  Either, then, all such scenarios are incorrect or the 
SUGAR directional determination is somewhat in error. 
If not the GC, then the SUGAR anisotropy could be due to the supernova remnant W28,
which has also been detected by the EGRET instrument in gamma rays.  However, W28 (located
at $\alpha=274^\circ$, $\delta=-23.18^\circ$) is itself displaced from the SUGAR position 
by about $4^\circ$, and one would need to again invoke an error in SUGAR's directional
determination.  For reasons we immediately explain, such an error seems to be a
viable possibility for, if the SUGAR directional determination is correct then
\begin{enumerate}
\item {\bf in the case that the anisotropy is due to neutrons} there must be a completely
unknown source at the position suggested by the SUGAR data (unobserved, e.g., by the
EGRET instrument in gamma rays)
\item or, alternatively, {\bf in the case that the anisotropy is due to protons directly}
either 
\begin{enumerate}
\item there is a source located very close to us which is
contrained to be in close {\it but completely coincidental} alignment with the
direction towards the GC 

or 

\item 
there is a source somewhat further away, the particles produced
by which -- {\it also completely coincidentally} -- happen to be bent in flight in exactly
such a way as to appear to come from close to the direction of the GC (and, further,
remain sufficently bunched that their signal is consistent with point-like for SUGAR).
\end{enumerate}
\end{enumerate}
 
We note, furthermore, that whereas, as stressed above, the GC is outside the field of view of 
AGASA, the position of the SUGAR maximum is {\it inside} the AGASA field of view so that the 
putative SUGAR source should be seen by AGASA. 
That it is not means that these two instruments 
are in disagreement. AGASA, moreover, has the better statistics.  We believe, then, that a 
quite natural reading of the situation is that SUGAR's directional determination is in error.
The only other way out of this apparent dilemma is to posit significant variability (between
the SUGAR and AGASA observation times) at the source which, however, 
would in fact be an argument 
in favor of a point source, rather than diffuse emission.


\section{Evidence for Hadronic Acceleration at the Galactic Center}

\label{section_GC_pi_decay}
Direct evidence of hadronic acceleration at the Galactic center comes from 
the EGRET detection of a $~30$ MeV - $10$ GeV continuum source 
(\3EG \ in the third EGRET catalog: Hartman {\em et~al.} (1999))
within $1^\circ$ of the nucleus (Mayer-Hasselwander et al. 1998)\footnote{The IBIS
telescope on board INTEGRAL has also recently released a preliminary result
for the detection of a GC source in the EGRET energy range: Di Cocco {\em et~al.} (2004)}.
The EGRET 
spectrum exhibits a clear break at $\sim 1$ GeV, and therefore cannot be fit by 
a single power law.  Instead, this break appears to be the signature of a process 
involving pion decays.  Specifically, the decay of neutral pions generated via
pp scatterings between relativistic and ambient protons produces a broad 
$\gamma$-ray feature that mirrors all but the lowest energy EGRET data.  Of 
course, pp scatterings also produce charged pions which, in turn, decay into
``secondary" electrons and positrons.  These leptons are capable of producing 
their own $\gamma$-ray emission via bremsstrahlung and Compton scattering.  
Interestingly, if the secondary leptons build up to a steady-state distribution 
balanced by bremsstrahlung and Coulomb losses, the former accounts naturally 
for the lowest energy EGRET datum, {\it independent of the ambient proton number 
density}.  This crucial feature results from the fact that the secondary leptons 
produce a steady state distribution whose normalization scales as the inverse of 
the ambient proton number density, whereas the bremsstrahlung emissivity per 
lepton scales directly with this density.  The pion decays link the lepton 
and photon generation rates, so the bremsstrahlung and pion-decay photon 
emissivities are tightly correlated.

While this discussion is quite general in nature, it is important to note that 
Sgr A East, a mixed-morphology SNR located within several parsecs of the Galactic 
center, is a viable candidate for the site of hadronic acceleration (see \S~7.2 
for a more complete discussion).  Specifically, the observation of OH maser emission
at the boundary of this structure provides strong evidence for the presence of shocks 
(Yusef-Zadeh et al. 1996, 1999).  In addition, Fatuzzo \& Melia (2003) have found that
a power-law distribution of shock-accelerated relativistic protons injected into the 
high-density, strongly magnetized Sgr A East enviroment leads to a pion-decay process 
(described above) that can account for both the EGRET source 3EG J1746-2851 
and the unique radio characteristics of Sgr A East. This scenario may also account 
for the $e^+ - e^-$ annihilation radiation observed from the galactic bulge by the 
Oriented Scintillation Spectrometer Experiment (OSSE) aboard the {\it Compton Gamma 
Ray Observatory} ({Fatuzzo}, {Melia}  and {Rafelski} 2001).

Finally, further evidence for the occurence of hadronic acceleration at the Galactic
center is presented by the detection of this region at $\sim$ TeV energies by
a number of air {\v C}erenkov telescopes: see \S \ref{section_TeV} for more on
this point.

\section{Relating a Pion-Decay \gam -ray flux to a Neutron Flux}

The processes via which an impinging, EHE charged beam can lead to the production
of astrophysical neutrons can be summarized as (with the `beam' particle
indicated first in each pair, the target second):
\bea
1.&\mathrm{pp}& \nn \\
2.&\mathrm{Ap},& \mathrm{A} \neq \mathrm{p} \nn \\
3.&\mathrm{p}\gamma& \nn \\
4.&\mathrm{A}\gamma,& \mathrm{A}  \neq \mathrm{p}. \nn 
\eea
In more detail, these processes are:
\begin{enumerate}
\item Leading neutron production from  collisions of accelerated
protons with ambient, target protons. 

\item Neutron production via dissociation of
accelerated
ions through collisions with ambient, target protons.

\item Photo-production of
leading neutrons (i.e., charge-exchange production of neutrons in collisions of accelerated
protons with ambient, target photons).

\item Photo-dissociation (fragmentation) of accelerated ions.
\end{enumerate}

Of the existing work concerned with the GC anisotropy and its explanation in 
terms of neutron production at various possible GC sites, the AGASA group's
anisotropy discovery paper \citep{Hayashida1999}
focused on the idea that the disintegration of accelerated
heavy ions
by interactions with ambient matter or photons was the ultimate source of the
(putative) EHE neutrons. This follows the lines of the broad scenario 
investigated by Sikora (1989)
for active galactic nuclei in which the very 
same strong
magnetic fields that serve to accelerate charged particles to high energies also serve
to confine the same to some central accelerating region (whereas neutrons escape).
Also of relevance here is the work of Tkaczyk (1994).

Alternatively to the heavy-ion disintegration idea, Medina-Tanco and Watson (2001) 
have proposed that a more likely method of EHE neutron production is via interactions 
between accelerated protons and ambient protons or IR \gam 's.
They found that the environment of Sgr A$^*$ -- the supermassive black hole at the
GC -- is sufficiently dense that the  particle interaction rate  required
to produce the desired neutron flux is achievable. We shall have more to say about this 
scenario below.

Takahashi and Nagataki (2001) also considered neutron production and determined that it is
pp interactions which most effectively produce the required neutrons. For reasons
we shall explain below, we agree with this conclusion (though our calculations differ importantly
in specifics). Takahashi and Nagataki (2001) also
researched the detectability of neutrinos concomittant with neutron production. More 
recently, Anchordoqui et al. (2003) have studied the detectability of
neutrinos produced in the decay-in-flight of the (putative) GC neutron beam and
Biermann et al (2004) have considered a model in which the observed anisotropy
is explained as due to the last GRB to go off in the Galaxy.

\subsection{Detailed Calculation of Neutron Flux from pp Collisions}

Protons accelerated to relativistic energies at the GC source
can undergo a series of interactions including 
$p N\rightarrow p N\, m_\mathrm{meson} \, m_{N\bar N}$,
where $N$ is either a $p$ or a neutron $n$, $m_\mathrm{meson}$ denotes the
energy-dependent multiplicity of mesons (mostly pions), and $m_{N\bar N}$ is the multiplicity of 
nucleon/anti-nucleon pairs (both increasing functions of energy).  
Since $m_{N\bar N}/m_\mathrm{meson} < 10^{-3}$ at low energy and even smaller
at higher energies \cite{Cline1988}, following Markoff et al. (1997) we here ignore
the anti-nucleon production events. The charge exchange interaction
($p\rightarrow n$) occurs around $40\%$ of the time at accelerator
energies and this fraction is predicted to be only very weakly energy-dependent (see the appendix).
We shall take it, then, that the {\it leading} neutron multiplicity, $m_n$, is given
by a fixed proportion of 0.4 (i.e., 40\% of all pp interactions involve
charge exchange, independent of incoming proton energy).

Other possible interactions of accelerated $p$'s -- all potentially important
for cooling -- are 
$p\gamma\rightarrow p\pi^0 \gamma$, $p\gamma\rightarrow n \pi^+ \gamma$,
$p \gamma\rightarrow e^+e^-p$ and $p e \rightarrow e N m_\mathrm{meson}$ (Markoff et al. 1997).

\subsubsection{The production of $\pi^0$ decay photons}

The (differential) $\pi^0$ emissivity resulting from an isotropic
distribution of shock accelerated protons 
$d n_p(E_p)/d E_p$ (where $[d n_p(E_p)/d E_p]$ is in units of cm$^{-3}$ eV$^{-1}$)
interacting with cold (fixed target) ambient hydrogen 
of density $n_H$ is given by the expression
\be
Q_{\pi^0}^{pp}(E_{\pi^0}) = c \; n_H \int_{E_p^{th}(E_{\pi^0})} 
dE_p\, 
\frac{d n_p(E_p)}{d E_p} \;
\frac{d\sigma(E_{\pi^0}, E_p)}{ dE_{\pi^0}}\;,
\ee
where $E_p^{th}(E_{\pi^0})$ is the minimum proton energy
required to produce a pion with total energy $E_{\pi^0}$
(determined through kinematical considerations) and $[Q_{\pi^0}^{pp}] =$
pions s$^{-1}$ cm$^{-3}$ eV$^{-1}$.
The resulting $\gamma$-ray emissivity is then
given by the expression
\be
Q_\gamma(E_\gamma) = 2 \int_{E_{\pi^0}^{min} (E_\gamma)}
dE_{\pi^0} \frac{Q_{\pi^0}^{pp}(E_{\pi^0})}{  (E_{\pi^0}^2 - 
m_{\pi^0}^2 )^{1/2}} \;,
\ee
where $E_{\pi^0}^{min} (E_\gamma) = E_\gamma + m_{\pi^0}^2  / (4E_\gamma)$.

At proton energies, $E_p$, 
 greater than $\sim$ 5 GeV---above the $\Delta$ resonance-affected region---the 
differential cross-section 
is approximated by the scaling form of Blasi \& Melia (2003; see also
Blasi \& Colafrancesco 1999):  
\be
\frac{d\sigma (E_p, E_{\pi^0})}{ dE_{\pi^0}} = \frac{\sigma_0}{ E_{\pi^0}} 
f_{\pi^0} (x_0)\;,
\ee
where $x_0 \equiv E_{\pi^0} / E_p$, $\sigma_0 = 32$ mbarn, 
and \citep{Hillas1980}
\be
f_{\pi^0} (x_0) = 0.67(1-x_0)^{3.5} + 0.5e^{-18x_0}\;.
\ee
This scaling form properly takes into account the high pion multiplicities
which occur at high energies. 

Given the above and a parent proton distribution
governed by a power law,  $d n_p(E_p)/d E_p$ with
spectral index $\gamma$, 
\be
\frac{d n_p(E_p)}{d E_p} \propto E_p^{-\gamma} \,
\label{eqn_power_law}
\ee
we can write the 
neutral pion emissivity due to pp as
\begin{equation}
Q_{\pi^0}^{pp}(E_{\pi^0}) = 
\frac{d n_p(E_{\pi^0})}{d E_p}\,\sigma_0\,n_Hc\,\Lambda^0(\gamma)\;,
\label{eq_pi_gamma}
\end{equation}
and, consequently, the
photon emissivity due to the decay of these
$\pi^0$'s as
\begin{equation}
Q_\gamma(E_\gamma) \simeq cn_H\,\int^\infty_{E_\gamma}\,
dE_{\pi^0}\,\int^\infty_{E_{\pi^0}}\,dE_p\; \frac{dn_p(E_p)}{ d E_p}\,
{d\sigma(E_p,E_{\pi^0})\over dE_{\pi^0}}\frac{2}{E_{\pi^0}}
\simeq
{2\over\gamma}\,
\frac{d n_p(E_\gamma)}{d E_p}\,\sigma_0\,n_Hc\,\Lambda^0(\gamma)\;,
\label{eq_gamma}
\end{equation}
where in both equations immediately above we employ
\begin{equation}
\Lambda^{0}(\gamma)\equiv\int^1_0dx_0\;x_0^{\gamma-2}\,f_{\pi^0}
= 2\left\{\Gamma(\gamma - 1)\left[18^{1-\gamma} 
+ \frac{15.5865}{\Gamma(3.5 + \gamma)} 
\right] - E(2-\gamma;18)\right\}\;,
\label{eq_lambda}
 \end{equation}
in which $\Gamma(x)$ is the Euler Gamma function and $E(n;z)$ is the
exponential integral function which satisfies $E(n;z) \equiv \int^\infty_1 \exp(-zt)/t^n 
\; \mathrm{d} t$.

\subsubsection{The Production of Neutrons in the Scaling Regime}

Similarly to the above, the emissivity of neutrons from an isotropic
distribution of shock accelerated protons $d n_p(E_p)/d E_p$ 
interacting with cold (fixed target) ambient hydrogen 
of density $n_H$ is given by the expression
\be
Q_{n}^{pp}(E_{n}) = c \; n_H \int_{E_p^{th}(E_{n})} 
dE_p\, \frac{d n_p(E_p)}{d E_p} \frac{d\sigma(E_{n}, E_p)}{dE_{n}}\;.
\ee
To proceed further we would like to take, in analogy to the above,
\be
\frac{d\sigma (E_p, E_{n})}{ dE_{n}} = \frac{\sigma_0}{ E_{n}} 
f_{n} (x_n)\;,
\label{eqn_neutron_diff_X_section}
\ee
where $x_n \equiv E_n/E_p$.
We must now determine an expression for
 $f_{n} (x_n)$.
In this regard, we employ the formalism set out in Appendix A
of Drury et al. (1994). Following Gaisser (1990), this 
reference sets out the calculation of
the `spectrum-weighted moment' (SWM), denoted by $\langle m x_S\rangle^\gamma_S$, 
for the emission spectrum
of various particle species
(labeled by $S$) 
from collisions of protons from a power-law distribution equation (\ref{eqn_power_law})
and where $x_S \equiv E_S/E_p$. 
In this formalism, the emissivity of species $S$ can be written
\be
Q_S(E_S) =  \frac{d n_p(E_S)}{d E_p} \sigma_{pp} n_H c \langle m x\rangle^\gamma_S,
\ee
so that we see that
$\Lambda^{0}(\gamma)$ defined in equation (\ref{eq_lambda}) above is nothing but the
SWM for neutral pions produced in pp collisions.

Drury et al. (1994) provide their own calculation of the
SWM for neutral pions, decay gammas, and neutrons
($\langle m x\rangle^\gamma_{\pi^0}$, $\langle m x\rangle^\gamma_{\gamma}$,
and $\langle m x\rangle^\gamma_n$).
We set out their results and ours (calculated with the $f_{\pi^0}$
given above) for comparison for pions and pion decay gammas in Table \ref{table_comps}.

To arrive at a SWM for neutron production calculations, 
Drury et al. (1994)
employ the 
dimensionless, inclusive cross-section for neutron production given by Jones (1990), viz
\be
g_n(x_n)= n_n(\alpha_n +1)(1-x_n)^{\alpha_n}.
\ee
Jones (1990)
gives, on the basis of his analysis of 300 GeV proton collider data, $\alpha_n = 2$,
and an average neutron multiplicity, $n_n$, of 0.25. 
We adopt his results excepting the neutron multiplicity
which we revise to be 0.4: see the appendix for more detail on this issue.
We find, then, that
\be
g_n(x_n) \to 1.2 (1-x_n)^2.
\ee
The translation of this distribution into our formalism is simply
\be
f_{n} (x_n) \equiv x_n g_n(x_n) = 1.2 x_n(1-x_n)^2.
\label{eqn_fn}
\ee
We find, then, 
\be
\label{eqn_neutemis}
Q_{n}^{pp}(E_{n}) \simeq c \; n_H \sigma_0 \frac{d n_p(E_n)}{d E_p} \Lambda^{n}(\gamma) \;,
\ee
where we have defined
\be
\Lambda^{n}(\gamma) \equiv\int^1_0dx_n\;x_n^{\gamma-2}\,f_{n}(x_n) \; = 
2.4 \;\frac{1}{\gamma(\gamma+1)(\gamma+2)}\;.
\ee
\begin{table}[h]
\begin{tabular}{|c|cccc|} 
\hline
$\gamma$ & 2.0 & 2.2& 2.4& 2.6\\
\hline
$\Lambda^{0}(\gamma)$ & 0.177 &  0.113 &  0.076 &0.053  \\
$\langle m x\rangle^\gamma_{\pi^0}$ & 0.17 & 0.092& 0.066 & 0.048 \\
\hline
$\Lambda^{n}(\gamma)$ & 0.177& 0.103& 0.063& 0.041 \\
$\langle m x\rangle^\gamma_n$ & 0.19 & 0.094 & 0.051 & 0.030\\ 
\hline
\end{tabular}
\caption{Values of $\Lambda(\gamma)$ for both neutral pions
and neutrons from pp collisions compared to the spectrum-weighted distributions
for same calculated in Drury et al. (1994). 
\label{table_comps}
}
\end{table}

\subsubsection{Relating Photon and Neutron Emissivity and Fluxes in the Scaling Regime}

Now, from Eqs.(\ref{eq_gamma}) and (\ref{eqn_neutemis})
we have that
\be
Q_{n}^{pp}(E_{n}) = 0.8 \gamma 
\frac{\Lambda^{n}(\gamma)}{\Lambda^{0}(\gamma)} 
Q_\gamma(E^0_\gamma)
\left(\frac{E_n}{E^0_\gamma} \right)^{-\gamma}\, ,
\label{eqn_neut_gamma}
\ee
where we have also used the fact that both daughter $\gamma$
and neutron spectra will be
governed by the same power law as the parent protons. We have, then, related the
neutron emissivity at some energy $E_{n}$ to the photon emissivity at 
a normalization energy $E^0_\gamma$ (assuming scaling holds). We present
$(0.8 \gamma \Lambda^{n}[\gamma])/\Lambda^{0}[\gamma]$ for various
values  of the spectral index $\gamma$ in Table \ref{table_ratios}.
The growth of this ratio with energy can be related to the fact that, because the average
energy of a neutron produced in a pp interaction (at lab energy $E_p$) will be
higher than the energy of a pion-decay \gam \ produced (indirectly) by a pp interaction
at the same energy ($E_p$), 
the neutron flux at $E_n$ is `directly' tied to the
photon flux at a somewhat lower energy. This means that as one steepens a spectrum
-- thereby increasing the number of photons in the population below some fixed
pivot point (at which the normalization is effected) -- one also tends to increase
the population of neutrons at $E_n$.


Note that 
in employing the relation set out in equation (\ref{eqn_neut_gamma}), careful attention
should be paid to the following points: 
\begin{enumerate}
\item
The neutron emissivity is
related not to the {\it total} photon emissivity but, rather, related to that
part of the photon signal due to pion decay. Determining an expectation, then,
for a neutron emissivity or flux requires that one be able to confidently
pin down what proportion of the \gam -ray signal at some normalization energy is due to
pion decay.
\item
In our derivation it was assumed that
$f_{\pi^0} (x)$ is given by its scaling form. This means that 
photon emissivity (or flux) data must be taken from
observations 
made at sufficiently high energy 
that we are guaranteed to be in the scaling regime, viz above $\sim$ 5 GeV.  
\item Likewise, it is assumed that the neutrons scale as indicated by
equation (\ref{eqn_fn}). This relation can be expected to go wrong at 
sufficiently high energies (see below for more on this point). 
Note that we are concerned with pp interactions at a lab energy of 
$\sim 5 \times 10^{18}$ eV entailing a cms energy range of
$\sqrt{s} \sim 70$ TeV.
\item
The parent proton power law will cut off at sufficiently
high energy. In making predictions for neutron
flux one must check that one is not above this cut-off energy.
\end{enumerate}

\begin{table}[h]
\begin{tabular}{|c|ccccc|} 
\hline
$\gamma$ & 2.0 & 2.1 & 2.2& 2.4& 2.6\\
\hline
$(\gamma \Lambda^{n}[\gamma])/(2 \Lambda^{0}[\gamma])$
&0.566&0.674&0.790&1.072&1.360\\
\hline
\end{tabular}
\caption{The ratio between photon and neutron
emissivities -- at the same energy -- from pp collisions in the scaling regime. 
\label{table_ratios}
}
\end{table}

\subsection{The Production of Neutrons at EHE: Accounting for Cross-Section Scaling Violation}
\label{sectn_Xsectngrth}

The above technology allows us to relate the neutron emissivity to the photon
emissivity of the same astrophysical object {\it provided that 
scaling holds}. That this caveat obtains can be seen directly 
from the fact that the integral of equation (\ref{eqn_neutron_diff_X_section})
over $E_n$ -- which should define the inclusive cross-section
for neutron production -- is, in fact, independent of center-of-mass
energy. This approximation holds good, by definition, over the scaling regime
from, say, 10 to 1000 GeV for the incident proton energy in the lab frame, 
but at the lab energies of over $10^9$ GeV that we are concerned with for EHE neutron 
production, it is no longer accurate (see Hagiwara {\em et~al.} 2002, fig. 39.12). To the level of accuracy 
required for the current application, however, it is not too difficult to 
account for the cross-section growth. To render the logic here most perspicuous, the
process can be described as a two-step one: (i) relate, via equation (\ref{eqn_neut_gamma}),
the photon and neutron emissivities at an energy scale small enough that
scaling holds and the cross-section can be taken
to be constant with respect to center-of-mass energy 
(ii) relate the high energy neutron emissivity to the low energy
neutron emissivity (inside the scaling regime) 
via the assumed power-law distribution and the ratios of the total pp 
cross-sections at these two energy regimes, $\sigma_{pp}(E_n)/\sigma_{pp}(E_\gamma)$.

\subsection{Neutron and $\gamma$-Ray Fluxes Related Given Neutron Decay}
\label{section_ngammafluxes}

Of course, we can equally well relate the observed pion decay
$\gamma$-ray flux at the
Earth to the
expected neutron flux (neglecting neutron decay -- and other effects that might
treat photons and neutrons differently in propagation -- for the moment).
Indeed, one may quickly determine that, given a power-law parent proton distribution,
and also accounting for the growth in the total cross-section discussed above,
\be
F_{n}^{n.d.}(E_{n}) \simeq 0.8 \, \gamma 
\frac{\Lambda^{n}(\gamma)}{\Lambda^{0}(\gamma)} F_\gamma(E^0_\gamma)
\left(\frac{E_n}{E^0_\gamma} \right)^{1-\gamma} 
\frac{\sigma_{pp}(\sim 10^{18} \, \textrm{eV})}{\sigma_{pp}(\sim 10^{12} \, \textrm{eV})}\, ,
\label{eqn_neut_flux}
\ee
where $F_{n}^{n.d.}(E_{n})$ denotes the total flux of neutrons
{\it above} $E_{n}$ that would be detected at Earth if the neutrons
did not decay ($n.d.$ denotes `no decay') and $F_\gamma(E^0_\gamma)$
denotes the total flux of $\pi$ decay photons detected above the normalization
energy
$E^0_\gamma$. Note that the above relation---directly relating
a neutron flux at Earth to a photon flux at Earth---obviates 
the need for an accurate estimate of the distance to the source. 
Also note  from fig. 39.12 of 
Hagiwara {\em et~al.} (2002)
that the ratio
of the pp total cross-sections pertinent to the $\gamma$-ray and
neutron production regimes is
\be
\frac{\sigma_{pp}(\sim 10^{18} \, \textrm{eV})}{\sigma_{pp}(\sim 10^{12} \, \textrm{eV})} 
\simeq \frac{150 \, \textrm{mb}}{40 \, \textrm{mb}} = 3.75 \, .
\label{eqn_Xsectn_ratio}
\ee

We are now in a position to provide a preliminary estimate
of the neutron flux above, say, $10^{17.9}$ eV based on the
EGRET data on the GC source. From the data provided in Mayer-Hasselwander et al. (1998),
the differential flux of photons at $\sim 6.3^{+3.7}_{-2.3} \times 10^{9}$ eV is
$6 \times 10^{-18}$ cm$^{-2}$ s $^{-1}$ eV$^{-1}$. With a power-law spectrum
of spectral index $\gamma$ this translates to a gamma ray flux 
\be
F_\gamma(6.3 \times 10^{9} \;\mathrm{eV}) = \frac{(3.8^{+3.2}_{-1.8})}
{\gamma - 1} \times 10^{-8}
\;\mathrm{cm}^{-2} \;\mathrm{s}^{-1}. 
\label{eqn_gamma_flux}
\ee 
Substituting this photon flux into equation (\ref{eqn_neut_flux})
we find the values for neutron flux tabulated against $\gamma$ in the 
first row of Table
\ref{table_neut_flux}.

As stressed, however,
these preliminary flux estimates do not take into account
propagation effects. In this regard, we can quickly dismiss any potential effect  from  
attenuation of the neutrons due to collisions with ambient particles in propagation:
here the greatest {\it potential} effect would be due to collisions of the neutrons with
ambient protons (see \S \ref{sectn_pgamma} below). Now, 
based on extinction at IR and optical wavelengths, 
the column density to the GC
is at most
around $5 \times 10^{22}$ cm$^{-2}$, implying a `grammage'
of matter that must be traversed by the GC neutrons in reaching the Earth of
around 0.1 g cm$^{-2}$ or less. Neutron energy losses
due to interaction with ambient matter in propagation
only become significant for grammages in excess of 80 g cm$^{-2}$ 
(Tkaczyk 1994),
however, and are, therefore, entirely negligible in this instance
(one can also establish that TeV photons from the GC
are not significantly attenuated by pair production in propagation: see \S \ref{section_TeV_attn}).

In contrast,
neutron decay must certainly be accounted for. We can incorporate
this effect by writing the flux as
\be
F_{n}(E_{n}) \simeq  
\int_{E_n}^\infty \frac{\mathrm{d} F_n^\mathrm{n.d.}(E_n)}{\mathrm{d} E_n}
\exp\left(-\frac{d_{GC}}{d_n[E_n]}\right)
\mathrm{d} E_n\, ,
 \label{eqn_neut_flux_adjusted}
\ee
where $d_{GC}$ is the distance to the Galactic Center, $d_n(E)$ is the neutron
mean free path defined in equation (\ref{eqn_mfp})
and the neutron differential flux without decay is given by 
(cf. equation \ref{eqn_neut_gamma})
\be
\frac{\mathrm{d} F_n^\mathrm{n.d.}(E_n)}{\mathrm{d} E_n} \simeq 3.0 \, \gamma 
\frac{\Lambda^{n}(\gamma)}{\Lambda^{0}(\gamma)} 
\frac{\mathrm{d} F_\gamma(E^0_\gamma)}{\mathrm{d} E^0_\gamma}
\left(\frac{E_n}{E^0_\gamma} \right)^{-\gamma}\, ,
\label{eqn_diff_flux_neut}
\ee
in which, as above, $E^0_\gamma$ is some normalizing energy at which the
$\pi^0$-decay photons are observed and the numerical factors from 
equations (\ref{eqn_neut_flux}) and (\ref{eqn_Xsectn_ratio}) 
combine to give $3.75 \times 0.8 = 3.0$.
Note that now the distance to the source enters into the calculation.
We present the result of calculating the neutron flux with decay in the 
second line of Table
\ref{table_neut_flux}.
\begin{table}[h]
\begin{tabular}{|c|cccc|}
\hline
$\gamma$ & 2.1 & 2.2 & 2.3& 2.4\\
\hline
(i)
& $(1.1^{+1.7}_{-0.7}) \times 10^{-16}$
& $(1.7^{+3.1}_{-1.1}) \times 10^{-17}$
& $(3.0^{+5.6}_{-1.9}) \times 10^{-18}$
& $(4.9^{+9.9}_{-3.2}) \times 10^{-19}$\\ 
(ii)
& $(6.1^{+9.9}_{-3.7}) \times 10^{-17}$ 
& $(9.8^{+17.3}_{-6.1}) \times 10^{-18}$
& $(1.6^{+3.0}_{-1.0}) \times 10^{-18}$
& $(2.6^{+5.2}_{-1.7}) \times 10^{-19}$\\
\hline
\end{tabular}
\caption{Values for the flux of neutrons above $10^{17.9}$ eV
due to the EGRET GC source 
for the cases where neutron decay-in-flight
(i) is {\it not} and (ii) {\it is} accounted for.
The units are neutrons cm$^{-2}$s$^{-1}$.
These numbers should be compared to the flux calculated `indirectly'
from the AGASA data by
\cite{Bossa2003}, viz. $2 \times 10^{-17}$ cm$^{-2}$ s$^{-1}$
 between
$10^{17.9}$ eV and $10^{18.5}$ eV,
and `directly' from the
SUGAR data for the observed point-like excess, viz
$(9 \pm 3 \times 10^{-18})$ cm$^{-2}$ s$^{-1}$
for the same energy range \citep{Bellido2001}.
\label{table_neut_flux}
}
\end{table}
As one can see from Table
\ref{table_neut_flux}, the putative
neutron source  seen (indirectly) by AGASA and (directly) by SUGAR is
easily accounted for by the GC EGRET source with a
spectral index of $\gamma = 2.2 \to 2.3$ (the best fit is at 2.23: see \S \ref{section_TeV_attn}). 
Not only is this range 
full compatible with the expectation from
theoretical calculations 
of shock acceleration spectra, it is also 
completely consistent with the value of 2.2
recently determined by Fatuzzo and Melia (2003) as their best fit for pion-decay \gam 's
to the EGRET source \3EG (see more on this point below) and, further,
is also consistent with the spectral index of 2.2 determined by Bossa et.~al (2003) in their
fit to the EHE AGASA data.
Note that other possible neutron production channels (besides pp collisions) are
discussed in \S \ref{sectn_alt_channels}.

\subsection{Detection of the Galactic Center at TeV Energies}
\label{section_TeV}

A Galactic center source has been detected by three air {\v C}erenkov telescopes (ACTs)
at $\cal{O}$[TeV] 
energies:
Whipple \citep{Kosack2004}, \cang \ \citep{Tsuchiya2004}, and, most recently and significantly,
HESS \citep{Aharonian2004}. In addition, the Hegra ACT instrument has put a (weak) upper limit on GC
emission at 4.5 GeV \citep{Aharonian2002} and the Milagro water 
{\v C}erenkov extensive air-shower array
has released a preliminary finding of a detection at similar energies from 
the `inner Galaxy' (defined as $l \in \{20^\circ, 100^\circ\}$ and $|b| < 5^\circ$: 
Fleysher (2002)).
Here
we address only the Whipple, CANGAROO, and HESS results in any detail, though note that
all the observations mentioned above lend crucial support to the notion that
acceleration of particles to very high energies is taking place at the GC.

Even restricting ourselves to consideration of results from these three instruments, we
find the situation somewhat confused regarding
the GC. In fact, it was clear,
even before the arrival of the recent, remarkable data from the HESS instrument, that the
Whipple and \cang \ GC observations were in conflict: Whipple has detected
(in data collected over a total of 26 hours from 1995 to 2003), at 
(conservatively) the
3.7 $\sigma$ level, a flux
of photons from the GC direction of
$
1.6 \pm 0.5 \; \mathrm{stat} \pm 0.3 \; \mathrm{sys} \times 10^{-12} \, \mathrm{cm}^{-2} \, \mathrm{s}^{-1}
\label{Whipple_flux}
$
above 2.8 TeV \citep{Kosack2004} which is
40\% of the Crab flux above this same energy. 
(In regards to the errors on the measuremnt,
note that there is a 20\% uncertainty, too, in the
callibration of the 2.8 TeV energy threshold). 
This is to be contrasted with GC data which were collected in 2001 and 2002 
by the \cang -II ACT. From these data
the \cang \  collaboration
has been able to generate a spectrum for the GC source in six energy bins.
From this spectrum can be extracted
\citep{Hooper2004} a flux of around
$
2 \times 10^{-10} \; \mathrm{cm}^{-2} \; \mathrm{s}^{-1}
$
above 250 GeV: this is at the level of 10\% of the Crab \citep{Tsuchiya2004}. 
The \cang \  Collaboration also determine an
extremely steep spectrum for their GC source with a fitted spectral index of $4.6 \pm 0.5$.

The fluxes of the Whipple and \cang \  sources (relative to the Crab)
are, then, quite different and a natural reading of the situation
is that the instruments are in conflict (see Hooper {\em et~al.} (2004) 
for further discussion on this point).
Note that any variability of the GC source at these energies
is now constrained to be small \citep{Kosack2004}
so that, though the two instruments in question
collected their GC data
over periods
largely non-coincident with each other, 
there is little leeway for explaining the difference between 
the two instruments by positing that they happened to  observe the source at different
activity levels.
Also note that the fields-of-view of the two instruments are similar
and that their respective GC sources are at similar positions and of similar extent
\footnote{Steve Fegan, private communication.}.

Adding very significantly to our knowledge of the GC at $\sim$ TeV energies,
the High Energy Stereoscopic System (HESS) Collaboration \citep{Hinton2004},
which employs four imaging atmospheric Cerenkov telescopes, 
has recently released TeV-range, GC data
unprecedented in its detail. This group has detected a signal in observations
conducted over two epochs (June/July 2003 and July/August 2003)
with a 6.1 $\sigma$ excess evident in the former and a 9.2 
$\sigma$ excess in the latter 
\citep{Aharonian2004}.
The data from the larger, July/August 2003 data set (which we shall use in our analysis)
can be fitted by a power law with, from the collaboration's own determination
\citep{Aharonian2004}, a spectral index 2.21 $\pm$ 0.09 and normalization
$(2.50 \pm 0.21) \times 10^{-8}$ m$^{-2}$ s$^{-1}$ TeV$^{-1}$ with a total flux above
the instrument's 165 GeV threshold of $(1.82 \pm 0.22) \times 10^{-7}$ m$^{-2}$ s$^{-1}$
(there is also a 15-20\% error from energy resolution uncertainty). 
 Data from the June/July 2003 run are consistent within errors
with the July/August 2003 data.

The HESS flux determination is
equivalent to 5\% of that from the Crab 
above the 165 GeV threshold. 
It is in conflict with the results from Whipple
 (see figures 1 and 2), the latter's flux determination being
a factor of three above that implied by the HESS spectrum \citep{Aharonian2004}.
It is also in
striking conflict with the
CANGAROO data: the HESS spectral index determination is clearly at variance
with the very steep spectrum
found by CANGAROO (see figure 4 of 
Aharonian (2004)
for a clear illustration of this
fact). Again, one might interpret these discrepancies as evidence for significant source
variation between the the instruments' different observing periods (i.e., over timescales of
$\sim$ year) but the multi-month
HESS data no more indicate source variability {\it within} the observing
period than the previous observations \citep{Aharonian2004}.

Intriguingly, the HESS data are also  difficult to reconcile with the EGRET GC data
(again, see figures 1 and 2 and also see the inset of figure 4 of Aharonian(2004))
-- if one assumes
a single source. We shall have more to say about this issue below. For the moment, however,
we must clearly decide upon which of the TeV data sets we should base our analysis.
Certainly one telling point  against the 
\cang \ data, as noted above, 
is the unusually steep, GC source spectrum. 
Although we do not pretend to any in-depth knowledge 
of the workings of the \cang \  analysis,
we do note that
observations of instrinsically bright regions
like the GC are, generically, affected by the problem that at lower energies putative events can be
`bumped over' a detector's threshold by noise, whereas at higher energies such a
mechanism would not be expected to be 
working\footnote{One way around this potential problem is to employ a `padding' procedure in
which artificial noise is fed into the data to account for the systematic
brightness differences between on-source and background observations. This is at the cost
of reducing the signal-to-noise ratio and, therefore, raising the threshold for
observations but without such a procedure the artifical steepening of a power law is difficult
to avoid.
Now,
the VERITAS Collaboration, in its analysis of the Whipple data from sources located in bright
regions employs exactly the padding procedure described immediately above 
(Steve Fegan, private communication). But, to the best of our knowledge,
the \cang \ collaboration does not use this procedure.
For this reason
we believe 
that the slope of the \cang \ data on the GC source are
questionable.}. A source spectrum, then, might
be made to seem steeper than in actuality by this
preferential recording of lower energy events. 
Furthermore, though the HESS and Whipple data are somewhat at variance,
they are certainly less in conflict than either result is with CANGAROO (again, see 
figure 4 of 
Aharonian (2004)),
so they support each other in a qualified sense.
For these reason we focus on the HESS and Whipple results and, then noting both the greater
detail and statistical weight in the former, we finally are drawn to the
conclusion that we should leave only the
HESS data in our analysis.

\section{Fitting to All Data}

We will now attempt to derive a best fit to the following
data:
the EGRET $\gamma$-ray  differential flux at $\sim$ GeV 
energies, the HESS $\sim$ TeV $\gamma$-ray flux 
and the 
EHE ($\sim$ $10^{18}$ eV) cosmic ray anisotropy data (assumed due to
neutrons). Such a fitting procedure makes sense in principle because, as shown above,
the neutron and photon fluxes are governed by power laws with the same spectral index 
(as both these species arise from the interactions of the same parent spectrum of accelerated protons) 
and at any energy we can relate the flux or differential flux of neutrons and photons.

Our procedure is to define a $\chi^2$ function which depends on the differences between 
fitted and
observed fluxes or diffential fluxes, weighted by the experimental error in each flux 
measurement
and also allowing for the systematic uncertainty in the energy calibration of the 
various instruments.
The free parameters in our analysis
are the spectral index, $\gamma$, and  the normalization of the $\gamma$-ray  
differential flux. 
We describe our procedure at greater length in an
appendix.



Employing this procedure, one quickly learns that the hypothesis
that the totality of \gam -ray and cosmic ray
data can be explained as arising from the interactions
of a single, parent population of shock-accelerated protons is not supported by the data:
the fit procedure produces a $\chi^2$ of 78.55 with a reduced $\chi^2$ of 6.54 for the
(14 - 2) degrees of freedom (dof) which is a very bad fit. There are a number of caveats 
to this bald assertion, however,
which we shall explore in somewhat greater detail below. These are, firstly, that 
it assumes that the detected radiation
-- in its various energy regimes -- is all equally unaffected in its propagation. A process which, say,
attenuated \gam -rays at TeV energies but not at GeV energies might be operating but this is not
accounted for in the fitting procedure. Secondly, 
the fits assumes that there is, effectively, only a single source. The GC is 
a highly energetic and dense region;
there could, in reality, then, be a
number of effective sources or acceleration/interaction regions (characterized by 
different magnetic field strengths, different shock compression ratios, different shock/magnetic field 
geometries, different ambient particle
densities, etc) there.
Thirdly,
the 
fit procedure assumes that the detected particles/radiation all originate in the same pp collision
process. If, say, another process were operating at high energies to supply some significant
fraction of the observed neutrons, then accounting for this process might allow for
 a better fit to the totality
of data (that said, the EGRET and HESS data are still difficult to reconcile).
We now explore these various points in greater detail.

\subsection{A Single Source: Attenuation of TeV Photons?}
\label{section_TeV_attn}

As should be expected from \S \ref{section_ngammafluxes},
if one performs a fit to only the EGRET+EHECR data
(`EHECR' denotes ~extremely high energy cosmic ray':
it is implicit in our analysis, of course, that these data are explained as arising
from neutron primaries), neglecting the  HESS data, one finds a very good fit: the total
$\chi^2$ is 0.33 for (4 - 2) degrees of freedom for a very plausible spectral index
of 2.23. We remind the reader that this value
is perfectly consistent with that determined by Fatuzzo and Melia (2003) for the
EGRET source \3EG \ and by Bossa, Mollerach, and Roulet (2003)
for the EHE CR source seen by AGASA (both these reference quote a figure of 2.2) and
is, moreover, perfectly consonant with the expectation from theory for acceleration at
strong shocks. Further,
the fitting points are around eight orders of magnitude apart so it is, indeed,
remarkable that the fit obtained between such widely-separated data points should require
a spectral index so close to the expectation from shock acceleration
theory and previous observations in more limited energy regimes (see the upper
line in figure 1).

\begin{figure*}[ht]
\epsfig{file=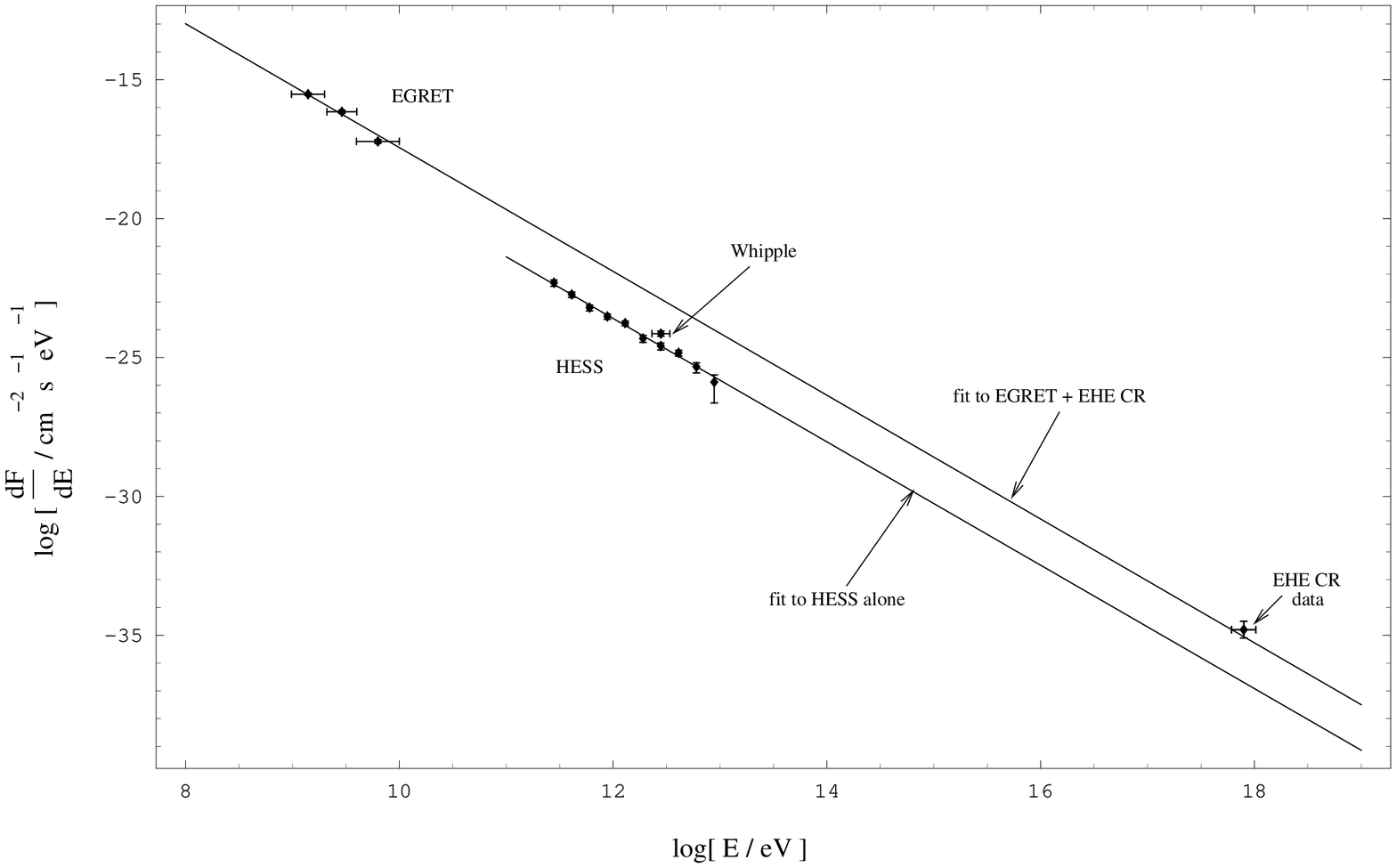,height=10cm,angle=0}
\caption{$\gamma$-ray and neutron differential fluxes together with fitted curves. 
The three points on the left of the figure are from EGRET (Mayer-Hasselwander et al. 1998). 
The 11 data points in the middle of the figure are GC, \gam -ray differential fluxes measured by
atmospheric cerenkov telescopes. Amongst these, 
the single point with marked error bars sitting proud of the 
fitted line is due to
Whipple \citep{Kosack2004}, the other
10 points are from the recent HESS July/August 2003 data set \citep{Aharonian2004}. 
The right data point gives the neutron flux which, on the basis of the
EHE cosmic ray data, we have taken to be 
 $1.0^{+1.0}_{-0.5} \times 10^{-17}$ cm$^{-2}$ s$^{-1}$
above
$10^{17.9}$ eV. 
The upper line gives the best-fit (as described in the text) photon differential flux
obtained from a 
simulataneous fit to the EGRET and EHE cosmic ray data. 
This is given by a power law
with a spectral index of 2.23 (the curve would be inaccurate at EHE because 
it does not take into account
the growth of the total pp cross-section). 
Obscured by (i.e., in excellent agreement with)
the right data point is a triangle indicating the position of
the neutron 
differential flux at $10^{17.9}$
eV as determined by the best-fit power law (that this point is apparently on top 
of the $\gamma$-ray flux
curve is coincidental). The lower curve -- with a spectral index of 2.22 -- has been found
by fitting a power law to the HESS data alone. Note how extremely closely the spectral indices
match.}
\label{fig_diff_flux}
\end{figure*}

The one problem with this scenario is the significant (by a factor of $\sim$20)
overprediction of TeV \gam -rays (again, see figure 1). We note that
there is no question but
that our fit to the EGRET+EHECR data does predict a TeV source
which is inside the field of view of
HESS's GC observations (and Whipple's for that matter)
and should have been seen by this instrument.  
The
simplest possibility to resolve the discrepancy, which we respectfully submit, 
is that the normalization of HESS GC
data is simply off by a large number. In support of this,
note that the other
remarkable aspect of figure 1 is that the lower line -- which is a fit to the
July/August 2003 HESS data alone --  so closely parallels the upper: it has a 
best-fit spectral index of 2.22\footnote{This spectral index has been
independently re-derived by us:
the HESS group find a spectral index of $2.21 \pm 0.09$ with a normalization of 
$2.5 10^{-8}$ m$^{-2}$ s$^{-1}$ TeV$^{-1}$ at 1 TeV. We perfectly agree with this
normalization to the limit of the significant figures quoted.}.
It does seem rather unlikely, however, that a normalization error could really
be as large as we require it and (given also that
even the larger Whipple flux determination is
still deficient with respect to our expectation from the
single source model) we are compelled to seek astrophysical explanations of the
data considering them to be fundamentally sound.

One  possibility, presaged above, is that some process is acting to attenuate
or downgrade the energies of the $\sim$ TeV photons {\it after} they have been generated, i.e.,
in their propagation from interaction point to us. In this regard, probably
the most attractive mechanism is pair-production on the optical-NIR background near the
source. Certainly something like this process is known to operate in the ``self-absorption''
of TeV \gam -rays from some X-ray binary systems by the thermal
photons emitted by those systems' own accretion disks (see Moskalenko (1995) for a review).
Pair production
 has an effective threshold which means that it does
not significantly attenuate \gam -rays with energies below a threshold, $E_\gamma^\mathrm{thresh(pair)}$,
given roughly by 
\be
E_\gamma^\mathrm{thresh(pair)} \sim  1 \; \left(\frac{E_\mathrm{bckgnd}}{0.5 \;\mathrm{eV} }\right)^{-1} 
\mathrm{TeV}\; ,
\label{eqn_pairprodn}										
\ee
where $E_\mathrm{bckgnd}$ is the typical energy of the (relevant) background photon population.
With NIR-optical background light, then,
the GC GeV signal would remain unattenuated as desired. From equation 
(\ref{eqn_pairprodn}) we require a background of $\sim$ 1.5 eV (i.e., NIR-optical) photons to attenuate
the HESS signal right down to the lowest datum at around 300 GeV.
Further considerations are the following:
\begin{enumerate}
\item Attenuation of the signal by 
$\sim 1/20$ of the expectation (given the fit to the other data) implies an optical
depth of $\ln(20) \sim 3$

\item  The peak cross-section for pair production is roughly
$10^{-25}$ cm$^2$, so we require a column density
of photons of around $3/10^{-25}$ cm$^2$ = $3 \times 10^{25}$ cm$^2$ to attenuate the photons. Over
the entire 8.5 kpc to the GC this would require an average optical-NIR photon number
density of $\sim$ 1000 cm$^3$. This is orders of magnitude
larger than what is found in the Galactic plane, so we shall concentrate on the idea that 
most attenuation will
happen very close to the GC source.

\item Given the similarities in the slopes of the fitted power laws
shown in figure 1,
the attenuation/degradation should be energy independent over the observed
TeV data points. At the heuristic level, at least, pair-production can
achieve something like this 
(though this is somewhat dependent on the distribution of the target photon population)
once one accounts for the possibility that the
daughter electron-positron pairs go on to produce further, high energy \gam -rays
by inverse Compton scattering of light in the background radiation field, 
thereby initiating a cascading process which redistributes the photon energy\footnote{In this regard,
consider
figures 6a and 6b of Carraminana (1992), which show the results of detailed
modeling of the ``self-attenuation'' of TeV \gam -rays from two X-ray
binary systems due to interactions with thermal NIR and optical photons emitted by those systems' own
accretion disks. It can be seen in these figures that the resultant (down-shifted) spectrum
parallels the unmodified spectrum for up to an order of magnitude in energy.}.

\item The main remaining question now is simply
whether we can plausibly get a sufficient column density of
NIR-optical photons near a candidate GC source to effect the attenuation/degradation.
For reasons that are explained in \S \ref{section_source_id} the two objects
we consider plausible sources for the
EHE cosmic rays are the accretion disk associated with the GC black hole itself,
Sgr A$^*$ and a supernova remnant located very close to the GC called Sgr A East. We consider the
photon environments in the immediate vicinity of each of these before briefly
considering the general, GC photon number density (i.e, within $\sim$ 10 pc of the center).

\begin{enumerate}
\item {\bf Sgr A$^*$}:
Genzel et~al. (2003) find that the ``local background
subtracted" luminosity of Sgr A$^*$ in the NIR is $\sim 3.8 \times 10^{34}$ erg.s$^{-1}$. 
This
means -- assuming a point source -- a column density along a radial
direction starting at $r_0$ is given by
$2.7 \times 10^{34} (r_0/\mathrm{cm})^{-1} $cm$^2$. Setting this equal to the required column density 
($3 \times 10^{25}$ cm$^2$) and
inverting we find $r_0 \sim 10^9$ cm. 
But this is inside the Schwarzschild radius
of the central black hole ($\sim 8 \times 10^{11}$ cm) and, therefore, an unphysical
requirement. The NIR light field due to the Sgr A$^*$ accretion disk, in other words, cannot attenuate
the TeV \gam -rays to the extent we require. We note in passing that
postulating a source location very close to the central black hole would present many 
observational difficulties. These are summarised in \S \ref{section_sgrastar}

\item  {\bf Sgr A East}: From fig 3 of Melia et~al. (1998)  the
extrapolated synchrotron flux at $\sim$ 1 eV for this object is 0.3 MeV cm$^{-2}$
s$^{-1}$ MeV$^{-1}$. This translates to a rough (number) luminosity of $2.6 \times 10^{45}$ 
photons s$^{-1}$,
which is an order of magnitude smaller than that for Sgr A*  which means an analogous $r_0$
many orders of magnitude too small given the $\sim$ pc scales of the Sgr A East shock(s).

\item {\bf General background}: On the other hand,
the actual quantity of interest is not
the background-subtracted
luminosity nor the luminosity due to any particular object. Rather, it is
the total number density of suitable photon targets in the GC
environment. This we can estimate from the following consideration:
Wolfire, Tielens and Hollenbach (1990) find that the GC circumnuclear disk requires an
ionizing UV photon flux of 100-1000 erg cm$^{-2}$ s$^{-1}$. 
Taking the upper figure
and assuming the same energy density in NIR photons\footnote{A re-processed IR photon background
of similar energy density to the 30 000 K UV
background is, in fact, expected, but this would realistically peak
at around 100 K $\sim 2.3 \times 10^{-2}$ eV: see \S \ref{sectn_pgamma}. Given that, 
in any case, 
we do not find any positive effect from the re-processed IR photons this
detail need not concern us.} one finds (at $\sim$1 eV) a
photon number density of $2.1 \times 10^4$ cm$^{-3}$.
Again, given the specified column density, this number density requires a
length scale of $1.5 \times 10^{21}$ cm $\sim 500$ pc to get sufficient attenuation. This
would seem to be excessive.
\end{enumerate}

\end{enumerate}

We reluctantly conclude, then, that around neither of the plausible, GC sources of the
EHE cosmic rays, nor in the general GC environment, does one find a large enough
NIR-optical photon number density (over sufficient scales) 
for our puposes: the optical depth to pair production experienced
by the
TeV \gam -rays
in their propagation out of the GC environment is too small for us
to explain the totality of data with a single source.
We now consider, therefore, the idea that two effective sources are 
contributing to the totality of data
with one source explaining the EGRET results and another (hopefully) able 
to account for both
the HESS and EHE cosmic ray observations. In this scenario we do not have 
a compelling explanation
for the closely parallel nature of the two fitted lines in figure 1 aside 
from the general expectation from
shock acceleration theory that the spectral index be close to 2.0 in a strong shock.

\subsection{Two Effective Sources?}

In introducing the idea that there may be two {\it effective} \footnote{We emphasise
`effective' here because two or more apparent sources might in fact orginate from 
a background population of protons accelerated in different regions of a
single, extended object.}
sources
we note that this is not an entirely unnatural reading of the situation.
There are two pieces of evidence we bring in here:
\begin{enumerate}
\item It has actually been determined by Hooper and Dingus (2002) in their
re-analysis of select data from the the 3EG catalog (Hartman et~al. 1999) that 
the GC is excluded at the 99.9 \% confidence
limit as the true position of the source
\3EG \ (this determination is at variance with the findings
made in the Hartman et~al. (1999) paper). Hooper and Dingus found the EGRET
source to be fairly well localised to a position 0.21$^\circ$
south west of the GC (i.e., $\sim$ 5 pc at this distance). 
In contrast, the HESS GC source was found to lie
with 95\% confidence within 0.05$^\circ$ of the GC \citep{Aharonian2004}.
\item
Both modeling and observation of SNRs over many years has consistently
shown that those examples with high flux 
tend to have a lower energy cutoff than those whose spectrum extends to 
higher energies.  This is generally attributed to the fact that physical 
conditions that sponsor efficient acceleration also lead to efficient 
cooling at higher energy. In general, then, 
the flux level and energy cutoff tend to go 
in opposite directions (see, e.g., Baring (1999)). 
\end{enumerate}

On the quantitative side,
the first item
we must now check is  whether a good fit is possible to the combined HESS+EHECR data
which, in this scenario, are supposed to be explained by a single source (we call this
assumed high-energy source 
the `HE source').
We find this, indeed, to be the case: such a fit
produces a $\chi^2$ value of 4.6 and a reduced $\chi^2$ of 0.51 for the (11 - 2) dof.
The best fit spectral index is very hard: 1.97. If we constrain the spectral
to be 2.0 and fit only to the normalization we obtain a $\chi^2$ of
5.0 and a reduced $\chi^2$ of 0.50 for (11 - 1) dof (see figure 2).

\begin{figure*}[ht]
\epsfig{file=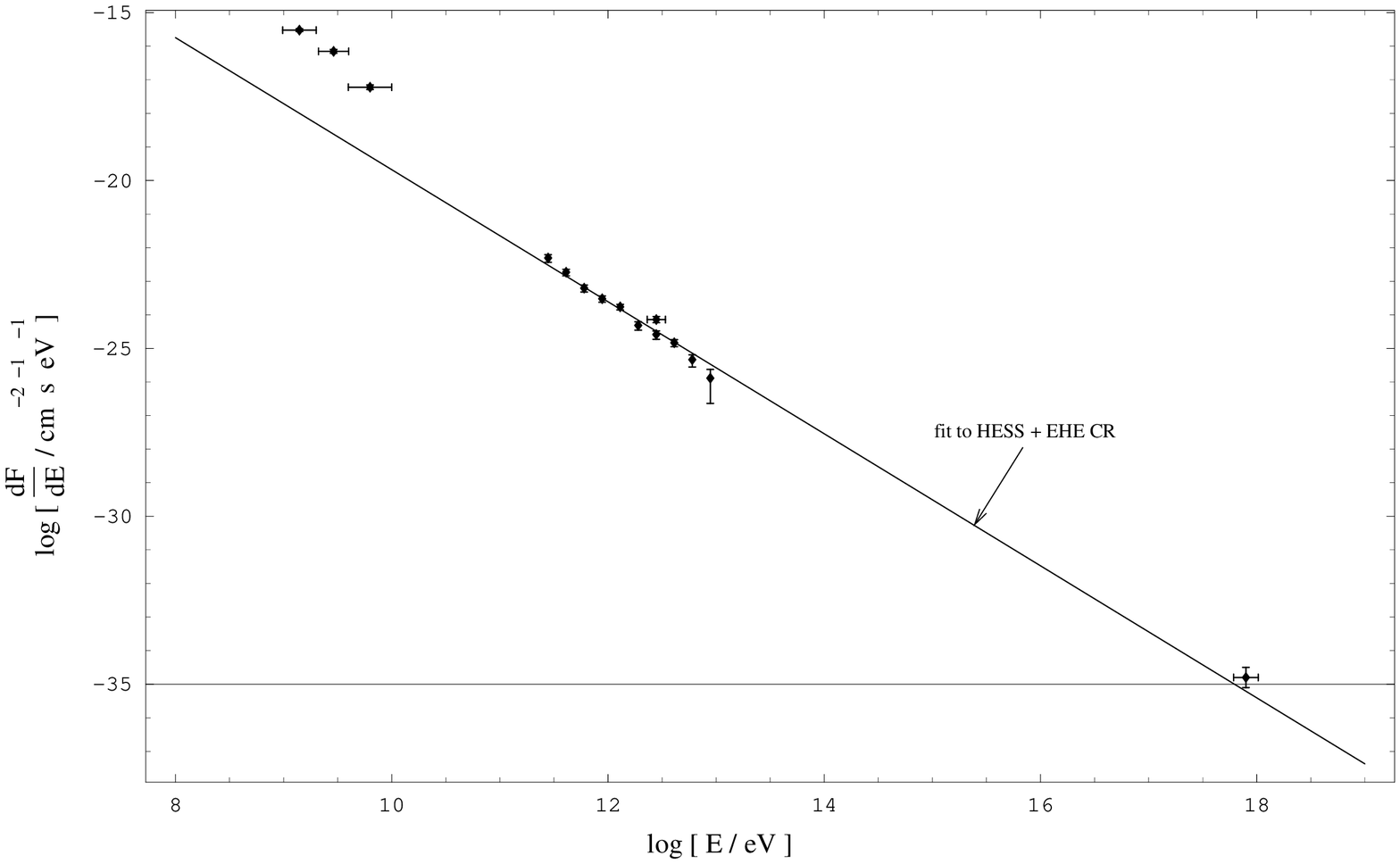,height=10cm,angle=0}
\caption{\gam -ray and neutron differential fluxes together with another fitted curve. 
The data are as given in the previous figure.
The curve is obtained from a power law fit (simulataneously) to the
HESS $\sim$ TeV \gam -ray data (but not the Whipple data point) 
and the EHE cosmic ray data
(again, note that the curve would be inaccurate at EHE because 
it does not take into account
the growth of the total pp cross-section). The best-fit spectral index is 1.97.
The EGRET data points have
{\it not} been used in this fit. 
On the right is illustrated both the EHE CR data point and (again obscured by the former)
a triangle indicating the position of
the neutron 
differential flux at $10^{17.9}$
eV as determined by the best-fit power law.}
\label{fig_diff_flux2}
\end{figure*}

The second issue we must confront is whether the other source (which we label
the `LE source')-- associated
with the signal seen by EGRET -- in any way interferes with the TeV observations which are
ascribed to the HE source. In particular, we must determine whether the LE source
is ``overtaken'' by the HE source at or below HESS energies as we require (in order that
$\sim$ TeV \gam -rays not be overproduced). This might happen in either (or in an effective
combination of) two ways: (i)
if the spectral slope
of the LE source is sufficiently steeper than the HE source and/or (ii) 
if the LE source cuts out below TeV energies
(the requirement that this source not produce a relatively significant flux 
of 300 GeV and above \gam -rays
would be guaranteed if the
parent protons cut out at or below $\sim$ TeV).

In regard to (i) and (ii) immediately above,
we note firstly that (i) appears not to be the case: we have tried two approaches
here  
and both {\it overpredict}
the differential \gam -ray fluxes at $\sim$ TeV by at least an order of magnitude. 
In the first approach
we fit only to the three (highest energy) EGRET data points for \3EG with variable
normalization and spectral index. This naive approach produces a very steep spectral 
index of 2.6. In the
second approach we fit only to normalization fixing the spectral index to 2.4 which 
is at the upper limit
of the spectral index range for the parent proton spectrum as determined by 
Fatuzzo and Melia (2003) in their fit to the totality of the EGRET data 
(i.e., all 9 data points).

We should consider, then, whether (ii) could describe the situation.

The interesting question now is, therefore, how we
arrange for the difference of something like seven orders of magnitude 
between 
the maximum 
energies attained in the LE and HE GC sources. Certainly variation in ambient particle densities
and magnetic field strengths will go some way towards explaining the difference, but it is
doubtful that the combination of these two could give a difference of $10^7$. 
Another tenable hypothesis is that the LE source -- which in this
scenario would be entirely independent of the HE source -- has an age-limited
maximum energy.
Another scenario would posit 
that the lion's share of the difference may be attributed to differences in shock 
geometries. In particular, one could 
postulate that the HE source realize a perpendicular 
shock configuration
and that the LE source be described by a parallel configuration. This difference
would be expected to contribute to at least two orders of magnitude variation in
maximum acceleration energies (all other considerations aside): see \S \ref{limitingE}, 
in particular,
equations (\ref{eqn_lag}) and 
(\ref{eqn_E_max_perp})
below. Note that we would also expect different shock compression ratios in the different 
effective sources
so the generic expectation would be for differing spectral indices.

\subsection{Other Neutron Production Channels}
\label{sectn_alt_channels}

There are two other channels which might reasonably contribute to EHE neutron
production at the GC, viz. p-\gam \ and heavy ion dissociation. Both of these operate,
effectively, without producing a concomitant GeV or TeV \gam -ray signal. This has the 
implication that our calculation of EHE neutron production (normalized
to this \gam -ray signal) is, all other things being
equal, a strict {\it under}-estimate.

\subsubsection{p$\gamma$ interactions}
\label{sectn_pgamma}

We have performed detailed calculations of the p\gam \ process
within the $\Delta$(1232) resonance approximation \cite{Stecker1979,Mucke1999,Dermer2002}. 
In this context,
the relevant neutron production channel is through the first
$\Delta$ resonance at 1232 GeV,
\be
p \gamma \to \Delta \to n \pi^+ \, ,
\ee
with a cross-section of about $6 \times 10^{-28}$ cm$^2$ \citep{Hagiwara2002}.
Here the branching ratios of the $\Delta$(1232) lead to
proton to neutron production in the well-defined ratio of
2:1 and its decay kinematics predict
a nucleon elasticity of 0.8 \citep{Mucke1999}. 
The interaction can take place if the energy of the ambient
photon in the p rest frame, $E_{\gamma(p)}^{thresh(pion)}$, satisfies
\be 
E_{\gamma(p)}^{thresh(pion)} \geq \frac{m_\Delta^2 - m_p^2}{2 m_p} \simeq 340 \, \textrm{MeV} \,.
\ee
This means that
the relevant target photon population in the GC context is supplied
by the intense flux of IR photons from the circumnuclear disk.
This is a powerful source ($\simeq 10^7 \lsun$) of re-processed
mid- to far-infrared continuum emission with a dust temperature
of $\simeq 100$ K
(Telesco 1996).
Given this temperature, we find that the p\gam \ process does not contribute
significantly for neutron production until proton energies of $\geq 6.9 \times 10^{18}$ eV
are reached, or, given the elasticity, does not contribute significantly
to neutrons with energies below $\sim 5.5 \times 10^{18}$ eV.
(The fact that this reaction does not `kick-in' until such high energies
explains why it
can not be directly normalized to the GeV \gam -ray signal: 
despite the fact that
the decay of neutral pions from the other branch of the $\Delta$ decay
will certainly produce  photons, these will all be directly
produced in the EHE regime.)
This fact -- our detailed modeling shows -- means that the p\gam \ process produces a EHE neutron
flux in the relevant energy range, at most,
$\cal{O}$[1\%] of that due to pp and may, therefore, be ignored.

Note that we have performed the calculations in this section assuming that the
interaction region is at $\cal{O}$[pc] scales from the GC (appropriate, e.g., to
the closer-in parts of the Sgr A East shell). We explain in \S \ref{section_sgrastar} why
it is unlikely that the interaction region be very much closer to the GC than this

\subsubsection{Heavy Ion Dissociation}

A calculation of
neutron production from dissociation of heavy ions 
(through interactions with either ambient
protons or light)
is beyond the scope of this paper 
(Tkaczyk 1994).
Such interactions, moreover, do not lead
directly to photon production at any energy and we can not, therefore, normalize
the rate of this process to the EGRET \gam -ray data directly. Still, that this
mechanism might be operating 
we find entirely plausible especially
given the fact that heavy ions make up a non-negligible
fraction of the detected cosmic ray population.

\section{Limiting Energies}
\label{limitingE}

As remarked above, the flux figures listed in Table \ref{table_neut_flux}
implicitly assume that the power-law description of the parent
protons which (indirectly) generate the photons observed at $\sim 5 \times 10^9$
eV continues
to hold up to {\it much} higher energies, $10^{18}$ eV and above.
It also assumes that the distribution of daughter neutrons
continues to be set by the scaling relation described in 
equation (\ref{eqn_fn}). 
This latter point we discuss in detail in an appendix. 
In brief,
we expect that the scaling relation will be correct to within a factor of 
two\footnote{Paolo Lipari, private communication, 2003.}, which, 
given other uncertainties in the calculation, does not introduce a significant
extra uncertainty in our neutron flux calculations.
Certainly, however, we need to devote
considerable attention to the former point regarding
the maximum energies to which any particular known source at the GC might accelerate
protons. 
We shall see that considerations surrounding limiting energy
actually mean that we can -- with some confidence -- identify which of the
potential sources at the GC is responsible both for the EGRET-observed signal and
the neutron signal.

In determing the limiting
energy, $E_p^\mathrm{max}$, two broad considerations play a part, viz (i) the
{\it intrinsic} limits to $E_p^\mathrm{max}$ given the macroscopic properties of the
shock at the source
doing the acceleration (magnetic field strength, shock size, geometry, and age, etc) and 
(ii) limits to $E_p$ from heating-cooling balance which we label
{\it scattering} limits.

We discuss both (i) and (ii) at greater length below.

\subsection{Perpendicular Shock Acceleration}

\subsubsection{Transport}

The whole of diffusive
shock acceleration theory can be obtained from the Parker equation, the fundamental
transport equation for the charged particle distribution function, $f(\vec{r}, t, p)$, 
in a background,
collisionless hydromagnetic field \citep{Parker1965}:
\begin{equation}
\frac{\partial f}{\partial t}
=\frac{\partial}{\partial x_i}\left[\kappa_{ij} \frac{\partial f}{\partial x_j}\right]
- U_i\frac{\partial f}{\partial x_i}
+ \frac{1}{3}\frac{\partial U_i}
{\partial x_i}\frac{\partial f}{\partial\ln (p)} + Q(x_i,t,p),
\end{equation}
where $\mathbf{\kappa}_{ij}$ is the
diffusion tensor and $Q(x_i,t,p)$ is the local source strength.
 This equation parametrizes the effects of
diffusion, convection, and acceleration/deceleration by an electric field on the
charged particle distribution. The equation assumes that the diffusion approximation
holds good, i.e., that particles are scattered often enough by magnetic field irregularities 
that the particle distribution is nearly isotropic. The Parker equation also requires that
the ratio of the shock speed, $U = |\mathbf{U}|$, to the particle speed, 
$w$ is small: $U/w \ll 1$.
In this picture, particle acceleration is caused by two factors (Jokipii 1982, 1987):
\begin{enumerate}
\item the large relative motion between the (magnetic)  scattering centers (i) causing the
diffusion in front
of the shock and (ii) behind the shock 
\item (if the magnetic field has a component {\it normal} to the direction of propagation
of the shock) drift along the shock front.
\end{enumerate}
Despite the potential importance of the second factor listed above, many discussions
neglect magnetic field changes and the resulting particle drifts, effectively
restricting consideration to quasi-parallel shocks.

From observation, it is known that the Galactic cosmic ray flux is quasi-isotropic to high
energy. This indicates that the Parker equation 
should hold as a valid description of cosmic ray acceleration. 
The scattering of a charged particle depends most strongly 
on the amplitude of magnetic turbulence 
on scales similar to that particle's gyroradius. This consideration justifies the
`quasi-linear' approximation in which the scattering rate is taken to be proportional
to the fluctuation spectrum of the magnetic field at a wavenumber approximately equal to the
inverse of the gyroradius.

\subsubsection{Acceleration}

Charged particle acceleration in the collisionless plasmas of space
is through the action of ambient electric fields, $\mathbf{E}$. Because the relevant
plasma processes are hydromagnetic (i.e., because we expect the charges to respond
to changes in the electromagnetic field on a time scale much shorter than the
flow time, so that the Lorentz force in the plasma frame is effectively zero) 
the electric field satisfies 
\be
\mathbf{E} = -\mathbf{U} \times \mathbf{B}. 
\ee
 Note that though $\mathbf{E}$ does not appear explicitly in the Parker equation, it is
implicitly contained in $\mathbf{U}$. We can arrive at a completely general
description of diffusive shock acceleration (i.e., applying equally well to
parallel and perpendicular shocks and everything in between) by putting a step-function
$\mathbf{U}$ into the Parker equation.

In the strong shock limit, $r$, the ratio between the upstream and 
downstream flow velocities satisfies 
\be
r = \frac{U_1}{U_2} \to 4
\ee
(note that a subsript of 1 indicates a quantity measured upstream or in front of the shock, 
whereas a subscript 2 indicates a quantity downstream or behind the shock). In this case, it can
be shown that  the 
steady-state solution of the Parker equation indicates that
the momentum dependence of the particle distribution 
goes like $f(p) \propto p^{-4}$ corresponding
to an energy spectrum $p^2 f \propto p^{-2}$. This describes
the universal power law (in either momentum or energy)
of spectral index close to 2.0 expected at strong shocks,
whatever the particulars of shock speed, diffusion coefficients, etc.

\subsubsection{Acceleration Rate}

In contrast to the spectral index, the high energy cut-off of the accelerated particles
{\it is} sensitive to the particulars of the shock. This maximum energy is
largely determined by the time available to accelerate the particles and the rate of energy gain.
These, in turn, are controlled by the finite lifetime of the shock itself, 
the escape of particles from the shock region, and
the rate at which particles scatter back
and forth across the shock (collision and synchrotron losses are also important, but these
are dealt with below).

By solving the time-dependent Parker equation  (and taking
particles to be injected into the shock at $t_0$ with momentum $p_0$),
we can determine 
that the particle distribution is still governed by the universal power law previously
determined but now with a high momentum cut-off, $p_c$, which satisfies
\be
\label{eqn_pc}
\frac{\textrm{d} p_c}{\textrm{d} t } \simeq U_\mathrm{shock}^2 \frac{ p_c}{ 4 \kappa_{xx}},
\ee
where $\kappa_{xx}$ is the diffusion coefficient normal to the shock front.
Either increasing the shock speed, $U_\mathrm{shock}$, or
 decreasing the diffusion coefficient, therefore,
will produce a greater rate of maximum momentum increase and, consequently, lead to 
faster particle acceleration in general.
Now, we can write the diffusion coefficient normal to the shock front as
\be
\kappa_{xx} = \kappa_\parallel \cos^2(\theta_B) + \kappa_\perp \sin^2(\theta_B),
\ee
where $\theta_B$ is the angle between the shock normal and the magnetic field vector.
One can see 
immediately, then, that, if the shock is quasi-perpendicular, 
$\kappa_{xx} \simeq \kappa_\perp$, because $\kappa_\perp$ is usually smaller than $\kappa_\parallel$,
such a shock will tend to accelerate particles faster than a quasi-parallel shock.

Let us consider the two limiting cases, quasi-parallel and quasi-perpendicular shocks, in a little
more detail.

\subsubsection{Quasi-parallel Shocks}

This is the more-commonly investigated case. From equation (\ref{eqn_pc}) 
we shall have that 
\be
\label{eqn_pll}
 \frac{1}{ p_c}\frac{\textrm{d} p_c}{\textrm{d} t } 
\simeq  \frac{ U_\mathrm{shock}^2}{4 \kappa_\parallel}
=  \frac{3 U_\mathrm{shock}^2}{4 \lambda_\parallel w},
\ee
where $\lambda_\parallel$ is the mean free path along the shock.
Now, given that this quantity must be of the order of or larger than the 
gyroradius of a particle at the limiting upper momentum, $r_g(p_c) \equiv r_c$,
we find that
\be
\label{eqn_Bohm}
 \frac{1}{ p_c}\frac{\textrm{d} p_c}{\textrm{d} t } 
\lesssim  
\frac{3 U_\mathrm{shock}^2}{4 r_c w},
\ee
which is the so-called Bohm limit on the acceleration rate. Many researchers have taken this to
represent a mechanism-independent limit on the acceleration rate and, hence, indirectly on
$E_\mathrm{max}$. The logic implicit here is that in order to be turned about so 
that it re-crosses the shock, 
a particle must be re-scattered each time it moves up- or downstream of the shock. The highest
energy gain rate, by this logic, will then occur for the {\it smallest} scattering mean free path,
which, in turn, cannot be smaller than the gyroradius.

\subsubsection{Quasi-perpendicular Shocks}

The reasoning presented above does not have universal validity, however: if the
magnetic field has a component perpendicular to the shock propagation direction and
if (as is usually the case as mentioned above) $\kappa_\parallel > \kappa_\perp$ (so that 
the particle is less constrained in motion {\it along} the shock than in motion away from it), the 
gyromotion of a charged particle can carry it across the shock many times between
each scattering. This can mean a much larger $E_\mathrm{max}$ than for the parallel case. 

To see this quantitatively, note that the kinetic theory (i.e., billiard ball scattering)
value for the ratio between perpendicular and parallel diffusion coefficients is given by
\be
\label{eqn_kap_rat}
\frac{\kappa_\perp}{\kappa_\parallel} = \frac{1}{1 + \left({\lambda_\parallel/r_g} \right)^2}, 
\ee
so that for a larger parallel mean free path, the perpendicular
diffusion coefficient is reduced and the acceleration is, therefore increased
(to be contrasted with the parallel case for which a smaller parallel mean free path
means an increase in the acceleration rate).

It seems that if we want acceleration to very high energies we should simply dial up
the required $\lambda_\parallel$.
This quantity may not be increased without limit, however. 
There will, in fact, be a maximum value for $\lambda_\parallel$ which can be 
determined (interchangeably) by the logic that (i)
the
diffusion approximation implicit in the Parker equation ceases to be valid
if particles do not scatter often enough to be isotropic at the shock or
(ii) particles must scatter often enough to diffuse upstream fast enough to stay ahead of the
shock 
\citep{Jokipii1982,Jokipii1987}
Either of these two conditions leads to the requirement that
\be
\label{eqn_Upsilon1}
\Upsilon \equiv U_\mathrm{shock} \frac{r_g}{\kappa_{\perp}} \ll 1,
\ee
or, for simple scattering
\be
\label{eqn_Upsilon2}
\Upsilon \simeq \frac{\lambda_\parallel}{r_g}\frac{U_\mathrm{shock}}{w}\ll 1,
\ee
so that if $U_\mathrm{shock}/w \ll 1$ then we can have $\lambda_\parallel/r_g \gg 1$
(while still maintaining Eq. \ref{eqn_Upsilon2}), giving us a significant
enhancement over the Bohm acceleration rate.

\subsubsection{Limiting Energies in Diffusive Shock Acceleration at Supernova Remnants}

Postulating that the maximum energy to which a  Supernova Remnant (SNR) shock
might accelerate particles, $E_\mathrm{max}^\mathrm{SNR}$ 
is given by (i) the Bohm value of the diffusion coefficient and
(ii) the modified Sedov solution for $U_\mathrm{shock}$,
Lagage et al. (1983) determined that
\be
\label{eqn_lag}
E_\mathrm{max}^\mathrm{SNR} \sim 
E_\mathrm{max}^\mathrm{L \& C}\equiv \mathrm{few} \times 10^{14} \times Z \mathrm{eV}
\ee
for the distance and magnetic field scales typical for a Galactic SNR.
Given that, as outlined above, 
SNRs are expected to accelerate the bulk of the cosmic rays up to at 
least the knee at $\sim 5 \times 10^{15}$ eV, equation (\ref{eqn_lag}) would seem to place a 
severe constraint on the mass composition of the cosmic rays. 

Following from 
the arguments presented above, however, Lagage and Cesarky's result will not
apply to quasi-perpendicular shocks. 
The sort of (globally) spherical shock produced by
a supernova, moreover, is quasi-perpendicular over much of its area, meaning that 
$\kappa_\perp$ is much more important that $\kappa_\parallel$ over a substantial fraction
of the SNR shock
and
the acceleration rate can be much larger than the
Bohm limit. This means $E_\mathrm{max}^\mathrm{L \& C}$ is actually
a significant {\it under}estimate of $E_\mathrm{max}^\mathrm{SNR}$ which, by
putting $\Upsilon$ equal to its maximum value, viz. $\sim 0.3$,  in equation 
(\ref{eqn_Upsilon2})
we can determine  to be
\be
\label{eqn_E_max_perp}
E_\mathrm{Z}^\mathrm{max(perp)} \sim 5 \times 10^{16} \times Z 
\left({B\over 20\;\mu G}\right)\left({R_{\mathrm shock}\over
10\;\mathrm{pc}}\right)
\mathrm{eV}
\ee
for a particle or nucleus of charge Z (see Appendix C for a full
derivation of the limiting energy). Note that in the equation
above ``perp'' signifies that the maximum energy appertains
to acceleration in a perpendicular shock geometry \citep{Jokipii1982,Jokipii1987}.
This energy is amply large enough to accelerate cosmic rays beyond the knee but note that
equation (\ref{eqn_E_max_perp}) 
represents an in-principle limit: in general, cooling effects 
due to synchrotron radiation
or collisions between the accelerated protons (and ions) and ambient matter or light may more
tightly limit the maximum energy/momentum to which the high energy particle population can be accelerated.
Further, the time available for acceleration is, of course, bound by the total age of the shock.
We now discuss these additional constraints.

\subsection{Cooling- and Age-Imposed Limits to Maximum Particle Energies}
\label{section_cooling}

There will, in general, be a 
maximum proton energy $E_p^\mathrm{max}$ above which  the combined energy loss
rate due to many processes -- proton synchrotron emission, inverse Compton scattering and
hadronic collisions  (with ambient nucleons and \gam 's) --
will exceed the rate of energy gain due to shock
acceleration.

In the GC environment, however, pp collisions are, by far, the dominant energy-loss
process (Markoff et al. 1997; Melia et al. 1998; Fatuzzo and Melia 2003).
This is the case becasue of the relatively lower density of target photons and the
intrinsically smaller cross-sections and inelasticities of $p\gamma$ processes in comparison with
pp collisions (this is to be contrasted with the situation that generally pertains in AGN, where
the photon density can be significantly higher than in the GC).

This maximum energy, $E_p^\mathrm{max(cool)}$, attainable given that pp collisions are
occuring,  depends 
on the functional form of the average collision inelasticity, where 
inelasticity denotes the fractional energy loss in the collision (i.e., the energy
of the leading baryon in units of the energy of the incoming proton).  
The time scale for proton cooling via pp collisions, then, is given by  
\be
t_{pp} \simeq \frac{1}{n_p \ c \,\sigma_{pp} K_{pp}},
\label{eqn_cool_time}
\ee
where the proton-proton cross-section, $\sigma_{pp}$, and
the fractional energy loss per pp collision or {\it inelasticity},
$K_{pp}$, are both, in principle dependent on energy (see the appendix). 

On the other hand, the proton (shock) acceleration time scale
is given by (Begelman 1990)
\be
t_p^\mathrm{acc}(E_p) = \frac{E_p}{\eta c^2 e B}\;,
\label{eqn_acc_time}
\ee
where $\eta$ is a dimensionless parameter, ${\cal O}$[1], that depends on the details of
the acceleration mechanism.

We can determine  $E_p^\mathrm{max}$, then, by setting
\be
t_{pp} = t_p^\mathrm{acc}(E_p^\mathrm{max})
\ee
 and inverting to find 
\be
E_p^\mathrm{max(cool)} = \frac{c \eta e B}{n_p \sigma_{pp} K_{pp}}.
\label{eqn_E_max_cool}
\ee
Finally in regard to maximum energies, one must also check that the time-scale
implied by equation (\ref{eqn_acc_time})
for the maximum energy calculated using equation (\ref{eqn_E_max_cool}) does not exceed
the age, $t_\mathrm{age}$, of the pertinent astrophysical object or environment. If this is the case, the
maximum energy will be age limited, i.e.,
\be
E_p^\mathrm{max(time)} = \eta c^2 e B t_\mathrm{age} \simeq 2.4 \times 10^{18} 
\left({B\over 20\;\mu G}\right)
\left({t_\mathrm{age}\over 20\;10^4 \; \mathrm{yr}}\right)
\mathrm{eV} \;  .
\label{eqn_E_max_time}
\ee
In general we shall have, then,
\be
\label{eqn_E_max_general}
E_p^\mathrm{max} = min \{E_p^\mathrm{max(perp)},E_p^\mathrm{max(cool)},E_p^\mathrm{max(time)}\},
\ee
where the maximum energies are given, respectively, by equations (\ref{eqn_E_max_perp}),
(\ref{eqn_E_max_cool}), and (\ref{eqn_E_max_time}).

\section{Identifying the GC Neutron Source}
\label{section_source_id}

\subsection{Sgr A$^*$ as EHE GC Neutron Source}
\label{section_sgrastar}

The question of the maximum energy to which protons might be accelerated at the
Sgr A$^*$ shock has been addressed at some length by 
Markoff et~al. (1999)
and here we
simply quote their results. 
As presaged above,
these authors found that pp collisions are the dominant energy loss process for relativistic
protons: $p\gamma$  processes are  supressed relative to 
the expectation from other galactic nuclei due to the very low luminosity of the GC and
pe$^-$ processes are also suppressed due to a dearth of extremely energetic electrons. 
Taking the mass of the central black hole to be
$2.61\pm0.65 \times 10^6 \msun$ \citep{Eckart1997}\footnote{Note that 
a recent revision to the Galactic center black hole mass  
that determines it to be $4.1 \pm 0.6 \msun$
\citep{Ghez2003}
does not substantially alter the conclusion of 
Markoff et~al. (1999)
regarding
maximum attainable energies for shocked protons.}, the shock to be located at
40 Schwarzschild radii, a (generously large) magnetic field of $\simeq 300$ G, 
and, finally, an ambient proton density of $\simeq 2 \times 10^8$ cm$^{-3}$, 
Markoff et al. (1999) determine that the maximum attainable Lorentz factor
for relativitic protons is approximately $4 \times 10^8$ which translates to 
$E_p^\mathrm{max} \simeq 4 \times 10^{17}$ eV. This maximum energy, then, means that
Sgr A$^*$ is probably not the source of the putative EHE neutrons seen by AGASA and SUGAR
(indeed, even if the anisotropy is due to the direct interactions of the primary,accelerated
protons, it still cannot
be the source). 

Another factor supporting the view that Sgr A* cannot be the source
of the observed EeV neutron flux would be the inconsistency implied by the emissivity
of secondary particles. The ensuing particle cascade would produce a copious supply of 
energetic electrons and positrons that, in the presence of the inferred $\sim 1-10$ Gauss 
magnetic field (let alone the $300$ G field we used for the estimate above) for this source, 
would greatly exceed Sgr A*'s observed radio flux.  This problem is alleviated if instead 
the secondary particles gyrate in the much weaker 
field associated with a typical supernova shell. Finally, 
Kosack et~al. (2004)
make the point that the lack of variability seen in their $\sim$ TeV
\gam -data for the GC source tells against the identification of 
this source with a compact point source such as Sgr A* (no evidence
for variability is seen in the EGRET data for \3EG \ either: 
Mayer-Hasselwander et~al. 1998).

\subsection{Sgr A East as EHE GC Neutron Source}
\label{section_sgraeast}






\subsubsection{Introduction to \sae}

\sae is a mixed-morphology, ${\cal O}[10^4]$-year-old
supernova remnant (SNR) located within 
several parsecs of the Galactic Center (GC) (see Melia and Falcke 2001).  
It appears to be of the class of 
SNRs which have been observed and detected at 1720 MHz (the transition energy of 
OH maser emission), such emission being a signature of shocks produced at the 
interface between supersonic outflow and the dense molecular cloud environments 
with which these SNRs are known to be interacting. There is, further, good 
evidence (see below) that  \sae \ may be associated with the EGRET source 
3EG~J1746-2851 introduced in \S \ref{section_GC_pi_decay}.  
It should be remarked, however, that if this putative association
is real, \sae \ has a \gam -ray luminosity ($\sim 2 \times 10^{37}$ erg s$^{-1}$) 
almost two orders of magnitude greater than that of the other EGRET-detected SNRs, 
something which clearly demands explanation.

In fact, in a self-consistent scenario explored elsewhere 
({Melia} et~al. 1998; {Fatuzzo} and {Melia} 2003)
the \sae \ remnant has 
been shown to be capable of producing the observed \gam -ray luminosity once the 
unusually-high ambient particle density ($\gtrsim 10^3 $ cm$^{-3}$) and strong 
magnetic field ($\gtrsim 0.1-0.2$ mG) at the GC are taken into account.  Indeed,
let us make the natural assumption that there is a population of protons  at 
\sae \ accelerated by the shocks associated with the supernova blast wave (and, 
therefore, governed by a power law with spectral index $\sim 2$). Then, from
collisions between these high energy protons and ambient ions, pions will be 
produced and these will, in turn, decay to produce \gam 's.  Now, the high-energy 
\gam -ray luminosity in this scenario is essentially dependent on the rate of
collisions between ambient ions and high energy protons, which is, in turn,
dependent on the product of the ambient ion density and the high energy proton 
density, $n_\mathrm{H} n_0$. So, we can fit to these two on the basis of the 
observed \gam -ray spectrum.  We also have, however, a handle on $n_\mathrm{H}$
from observations by ACIS on board the {\it Chandra X-ray Observatory}  -- \sae 
\ is expanding into an ionized gas halo with a density $\sim 10^3$ cm$^{-3}$
and interacting with  a cloud of density $\sim 10^5$ cm$^{-3}$. Further,
the total energy in the accelerated proton population is readily calculated
from $n_0$ and is, of course, constrained to be less than the total
energy released in the original explosion that gave birth to the \sae \ shell.

These considerations result in fairly rigorous contraints on any attempted fit 
to observations of \sae . But, with a proper treatment of pion production (in 
particular, the energy dependence of the pion multiplicity), the entire broadband
spectrum of the object can be compellingly explained. This explanation essentially 
sees the super-100 MeV \gam -ray spectum as due to $\pi^0$ decay, the sub-100 MeV 
\gam -ray spectum explained by bremsstrahlung emission (self-consistently with the 
assumed value for $n_\mathrm{H}$, namely $10^{3}$  cm$^{-3}$: see below) and the 
VLA observations of \sae \ at 6 and 20 cm explained by 
synchrotron emission (though see below for a caveat regarding the radio spectrum). 
Note that in this scenario, it is the secondary, charged leptons (ultimately 
resulting from collisions of the shock accelerated protons with ambient ions) 
that are responsible for the bremsstrahlung and synchrotron emission, rather than 
directly accelerated leptons. Indeed, the population of this primary lepton class 
can, in fact, be shown to be negligible in comparison with the secondaries, given 
that each accelerated proton produces of order 20-30 charged leptons 
\cite{Fatuzzo2003}.

In regard to Sgr A East's radio spectrum, it should be noted that
the synchrotron emission from the total population of secondary,
charged leptons is not {\it directly} consistent with radio observations:
the observed  spectral index is $\sim 1$, pointing to an underlying
population of non-thermal charged leptons with a power-law distribution of
index $\sim 3$ \citep{Pedlar1989}. In contrast, 
from general considerations (whether the lepton population is directly
accelerated or born of the interactions of shock accelerated protons),
we expect that the lepton index be $\sim 2$, leading to radio
emission characterized by a spectral index $\sim0.5$.
If, however, one allows for the reasonable eventuality that
high energy leptons diffuse out of the \sae \ region, the
convolution of a diffusion loss factor with
the initial lepton spectrum resulting from pion decay
allows a very good fit to the radio spectrum \citep{Fatuzzo2003}.
Alternatively, it might be that the high energy protons themselves
are diffusing out of the system thus also (indirectly)
depleting the high-energy
lepton spectrum. This scenario also provides a credible, though less good,
fit to the radio data. As will become clear, however, the very fact that
we do see a GC EHE cosmic-ray anisotropy argues against the latter scenario.

The scenario presented above -- which we label `HIGH' -- is to be contrasted 
with an alternative picture 
(`LOW') explored earlier \citep{Melia1998}, in which a considerably lower
ambient particle density was employed (viz. $\sim 10$ cm$^{-3}$). 
Through
the $n_\mathrm{H} n_0$ dependence of the \gam -ray luminosity this then
necessitated a large population of high energy protons, in turn
yielding an energy content of relatitivistic particles
of $\sim 8 \times 10^{50}$ erg. 
In contrast, the particle energy calculated for the HIGH scenario
is $\sim 10^{49}$ erg so
clearly the view that \sae \ resulted from a standard supernova
favors HIGH. Moreover, the recent {\it Chandra} observations of
\sae alluded to above 
support the view that \sae \ {\it is}, indeed, the remnant
of a single supernova with typical energetics.

\subsubsection{Maximum Energy of Protons Accelerated by the \sae Shock}

Typically, the magnetic field strength in the interstellar medium scales as the
square root of the density. In the case of Sgr A East, it appears that most of 
the pp scatterings (and essentially all of the bremsstrahlung and synchrotron
emissivity) occur in the $\sim 10^3$ cm$^{-3}$ region, where the field strength
is $\sim 0.2$ mG. Thus, if the shock occurs in the $\sim 10^5$ cm$^{-3}$ molecular
cloud colliding with the expanding shell, the field intensity within the particle
acceleration region may be as high as a few mG. However, a value greater than this 
seems implausible.

With a field strength of 4 mG, the other inputs described above, and the application of 
equation (\ref{eqn_E_max_perp}), one finds that particles in the Sgr A East shock can be 
accelerated, in principle, to $10^{19} R_{\rm shock}/10$ pc Z eV.  This figure
is not limited by cooling or age considerations: from equation (\ref{eqn_E_max_cool}) we find
that, given the assumed 3,000 cm$^{-3}$ ambient proton 
density\footnote{As is the case at Sgr A$^*$, pp collisions are the dominant cooling mechanism for
relativistic protons accelerated at the \sae shock \citep{Melia1998,Fatuzzo2003}.}, a lower limit on the
cooling-limited maximum energy is $2.7\,\eta\;10^{21}$ eV (this is a lower limit because
it has been derived assuming a pp cross-section of 150 mb, which, in fact, only pertains
at the very highest region of the pertinent energy range). Moreover, the time
required by the 4 mG field to accelerate protons to this energy is around 2 500 years, comfortably inside 
the  ${\cal O}[10^4]$ year age of the remnant (the age-limited energy using equation 
(\ref{eqn_E_max_time}) is around $\eta 5 \times 10^{20}$ eV for a 4 mG field and an age of
10 000 years).

Parenthetically, note from the above that
heavier ions might be accelerated to energies approaching
$10^{20}$ at \sae -- potentially explaining most of the cosmic ray spectrum even above the ankle.
If this were the case, it would {\it not} imply an unambiguous 
anisotropy towards the GC:
even at $10^{19}$ eV modeling shows that a proton leaving the vicinity of the GC would be
scattered many times, on average, before it could reach us at Earth. This mechanism
would isotropise the GC signal even at these extreme energies, 
though they would probably not completely wash it out. Such a scenario is not
ruled out by current observations.

\subsubsection{\sae and the Two Source Model}

We note finally that there being two effective sources required to 
explain the totality of data fits naturally into the hypothesis
that the Sgr A East shell accelerates 
all
the protons required to explain the various signals.
The Sgr A East shell ranges in distance from very close to the GC (within 1 pc)
to up to $\sim$ 10 pc. It subtends, therefore, regions of widely varying
ambient particle density and also magnetic field strength and orientation so that
proton populations accelerated in it would be expected to display or experience
varying
maximum energies, spectral indices and overall (energy) differential number
densities and interaction rates. Moreover, Hooper and Dingus's (2002)
determination that the EGRET source can be localized to a region 0.21$^\circ$ 
(i.e., $\sim$ 5 pc) away from the actual GC (while the HESS source is localized to
{\it within} 1 pc of the GC) supports this basic idea: the lower energy source
is located in a region where ambient magnetic fields can be expected to be lower and
also, perhaps, where ambient particle densities be higher, whereas the higher energy source
is located in a region of larger magnetic field strength. We hypothesise also that the
acceleration in these regions be described by different shock/magnetic field geometries (with
the EGRET source describing a parallel shock-magnetic field configuration and the
HESS/cosmic ray source being described by the more efficiently-accelerating perpendicular configuration).
Further observations and modeling will be required to determine whether this picture is 
 tenable.

\section{Conclusion}

We have examined -- and shown to be quantitatively successful --
the idea that the Galactic center 
EGRET source \3EG \ 
is also responsible for the observed anisotropy that is seen
in cosmic rays towards
the GC at an energy of around $10^{18}$ eV. 
We have, in particular, demonstrated that
a power-law fit to the $\sim$ GeV EGRET data and
the EHE cosmic ray anisotropy data (taking the latter to be evidence for
neutron primary particles)
requires a parent proton population with
a spectral index of 2.23. This is in exact agreement with
previous determinations made (individually) for the
spectral indices of the parent proton population that is generating the
EGRET-observed \gam -rays and for the distribution of the 
EHE cosmic ray signal. The determined spectral index is
also in perfect agreement with the expectation from shock acceleration theory.
Our fitting procedure requires input from the particle physics of charge-exchange in
pp collisions. Our results imply the existence of a population of shock-accelerated
protons (and perhaps heavier ions) at the GC with energies that range up to, at least, 
$\sim 10^{19}$ eV. 

Unfortunately, however, this fit overpredicts the flux of \gam -rays from the GC at $\sim$ TeV
energies given the existing data from a number
of atmospheric Cerenkov telescopes (which are, admittedly, somewhat at
variance amongst themselves). We show that what we believe to be the
best and most recent data -- that coming from the HESS instrument -- is well
fitted by a power law with a spectral index of 2.22, i.e., very close to the power law
fit to the combined EGRET and EHE cosmic ray data. One explanation, then,
of the deficiency of the $\sim$ TeV data with respect to expectation is that
the HESS instrument's normalization is wrong by a large (energy-independent) factor ($\sim$ 20).
Finding this somewhat unlikely, however, we have sought astrophysical
explanations of the totality of EGRET + HESS + EHECR data, taking these all to be
fundamentally sound. Pursuing the idea that all the data might be explained as
originating from the interactions of
a single high energy proton population, we searched for mechanisms that would (preferentially)
attenuate the $\sim$ TeV \gam -rays. The most promising candidate in this
direction, however, viz, pair production on an assumed 1.5 eV light background,
we have found is unlikely to be efficacious as the number density of such
photons in and around 
candidate GC sources or source regions is probably too low.

Noting these facts, we sought
for a model involving two effective sources. We showed here that 
the hypothesis that the EGRET data are explained by one source and the HESS + EHECR data
by another is perfectly tenable with, in particular, a good fit possible to the combined
HESS + EHECR data. This two source scenario -- which has some plausibility given
the varying source position determinations made on the basis of observations conducted in
the various energy regimes --
requires that the EGRET source probably be steeper than -- and certainly 
have a cut-off energy much lower than -- the HESS + EHECR source.

This difference in maximum particle energies could be attributed to 
either completely independent  populations of high energy
protons, accelerated in different astrophysical sources with
differing ages, magnetic field strengths, ambient particle densities, etc.
Alternatively -- and more plausibly in our opinion -- 
a population of high energy protons accelerated in different
regions (with differing magnetic field/shock geometries) 
of a single, extended object might
produce two (or more) effective sources.


We have shown that the 
 Sgr A East supernova remnant (located near the GC)
could provide the environment in which the acceleration of the
high energy protons we require could plausibly be achieved.
Further, we have shown that the `two-source' model fits naturally
with the hypothesis that Sgr A East is the accelerated proton source:
the Sgr A East shell ranges in distance from very close to the GC (within 1 pc)
to up to 10 pc distance. It subtends, therefore, regions of widely varying
ambient particle density and magnetic field strength and orientation so that
proton populations accelerated in it would be expected to display or experience
varying
maximum energies, spectral indices and overall (energy) differential number
densities and interaction rates. 
On the other hand, it seems that the other plausible GC source, namely the 
accretion disk around the GC black hole, Sgr A$^*$,
is unlikely to provide an environment in which acceleration of protons to the extreme
energies required is possible. Further, observations tell against
a point-like source for both the GeV and TeV regime \gam -rays.
Our results, then, provide tentative
evidence for the first direct connection between
a particular, Galactic object (Sgr A East)
and the acceleration of cosmic rays up to energies near the
ankle in the cosmic ray spectrum. 

At all energies, we anticipate -- with great interest -- the results
that shall be obtained by
the next generation of instruments. These instruments include the GLAST Large Area Telescope 
\citep{Thompson2004}
at $\sim$ GeV energies
and VERITAS \citep{Krennrich2004}, at $\sim$ TeV energies,
and, finally, the Pierre Auger Observatory \citep{Camin2004}
for cosmic rays of $10^{18}$ eV and higher. 
The results from these instruments will either
confirm or rule-out our scenario.

\section{Acknowledgments}

RMC gratefully acknowledges useful communications from Paolo Lipari
and enlightening conversations with Matt Holman, Jasmina Lazendic, Avi Loeb, and Josh Winn.
FM would like to thank Steve Fegan and Trevor Weekes for very helpful correspondence.
This research was supported in part by NASA grant NAG5-9205 and NSF grant 0402502 at
the University of Arizona. RRV and FM are also partially supported by a joint
Linkage-International grant from the Australian Research Council.
RRV thanks Paolo Lipari and Todor Stanev for general discussions
on cosmic rays and the observed anisotropies, and he thanks
Paolo Lipari and Maxim Khlopov for their kind hospitality at the
Universit\'{a} di Roma ``La Sapienza" where these discussions took
place. MF is supported by the Hauck Foundation through Xavier University.

\section{Appendix: Neutron Production in UHE Hadronic Interactions}


We discuss here the necessary, particle physics inputs to our calculation
of neutron production, viz. the 
total (inelastic) pp cross-section, the leading neutron multiplicity in pp
interactions and the leading neutron elasticity in pp.
In summary of the below, all three of these quantities might in principle
exhibit a scaling-violating dependence on the center of mass energy.
The data on the EHE cross-section are fairly sound and do indicate exactly such
a dependence, albeit a weak one. We account for cross-section growth as set out
in \S \ref{sectn_Xsectngrth}.
There is no data at EHE on neutron multiplicity or inelasticity
and, indeed, even at moderate energies the data are scant. Modeling,
however, predicts little or no energy dependence in these quantities
(note that we are talking about the production of
{\it leading} particles, not secondaries). We take it, then, that
the leading neutron multiplicity and elasticity
do not evolve from the values supplied by the low energy data.

\subsection{Leading Neutrons from pp Collisions: General Remarks}

By definition, leading  baryons (proton, neutron, or hyperon) 
created in pp (or pA) collisions
are those particles that
carry a large fraction of the incoming
hadron's momentum. 
On the other hand, these particles, can be expected to have, in general, a
low transverse momentum (see Eq. \ref{eqn_perp} below), i.e., 
leading particle production is a {\it soft} process. The transverse momentum transfer,
moreover, can almost be taken to be unrelated to the
energies of the primaries (or, indeed, the secondaries) and is, instead,
at an intuitive level at least, related to the inverse of the
`size' of the interacting hadrons $\sim 1/fm$ (e.g., Perkins (1987)).

To calculate the leading
neutron production rate in pp collisions with `beam' protons
of lab energy $\gtrsim 5 \times 10^{18}$ eV we need to know 
\begin{enumerate}
\item What the total (inelastic)
pp cross-section is at such an energy. 
\item  What proportion
of pp interactions at such energies result in the production of
{\it leading} neutrons, i.e.~the leading neutron {\it multiplicity}. We 
expect here a number less than unity that we could equally well
describe as the charge exchange probability (in pp interactions).
\item  What fraction of the incoming
proton's energy any leading neutron will typically have (i.e., the quantity
for the leading neutron that has traditionally been labeled the
{\it elasticity} by cosmic ray investigators).
\end{enumerate}
We discuss the empirical inputs to each of these three in turn below, before
we go on to consider theory and modeling.

\subsection{Hadronic Cross-sections at High Energies: Experiment}

Unlike the cases presented by leading particle multiplicities and inelasticities, 
hadronic cross-sections have been measured (almost) up to the highest
center-of-mass energies we are interested in ($\sqrt s \sim 50$ TeV),
though it should be noted
that at the highest energies the pp
cross-section is extracted -- using Glauber theory \citep{Glauber1970} --
from cosmic-ray data on proton-`air' nucleus interactions (see, e.g., \citet{Block2000}).
Moreover, between  $\sim 10^{11}$ eV and $10^{13}$ eV the cross-section
has not been directly measured. These two facts, however, do not
allow for any great uncertainty in the cross-section:
clear of the various resonances at a few GeV, the growth of the total
pp cross-section 
becomes smooth and slow in the measured region. Further,
there is a strict bound, the Froissart-Martin bound
\citep{Froissart1961,Martin1965}, on the asymptotic
behaviour of the total cross-section, viz., for $s \to \infty$ 
\be
\sigma_{tot} \leq C \ln^2 s.
\ee
From parametrization of the growth
of the total pp cross-section between
collider energies and 
UHE cosmic ray data, the Froissart bound may be saturated. 

As noted in the body of the paper, we take the pp cross-section to be
constant for beam energies 
between about $10^{10}$ eV and $10^{12}$ eV which, from
from fig. 39.12 of Hagiwara {\em et~al.} (2002) 
is quite reasonable. On the other hand, we do allow for the slow growth (Hagiwara 2002) of the
cross-section between this energy regime and that relevant to the
EHE neutron production.

\subsection{Leading Neutron Multiplicity in Proton-Proton Interactions: Experiment}

At the EeV energy regime we are interested in,
there is no direct, empirical data on leading neutron production. 
Indeed, there are, in general, great difficulties in obtaining information on production
of (even) charged leading particles from colliding beam devices (let alone neutrons).
This is because
the combination of large longitudinal and relatively
low transverse momentum  that, as mentioned above,
characterizes leading particles means that they are kinematically
constrained to emerge from their creation region at small angles. They
will, therefore, be traveling down a collider's beam pipe where there is little or no detector
acceptance.
In fact,
for $pp \to$ leading $n$, the most recent and highest energy data were obtained
in the mid 1970s
at experiments conducted
at up to $\sqrt s \sim 63$ GeV at the ISR \citep{Engler1975,Flauger1976} --- three orders of magnitude below the 
energy regime we are interested in ---
and Fermilab \citep{Dao1974}. These data are described to first order
by Feynman scaling \citep{Feynman1969}
which reflects the partonic make-up of the colliding protons (see below).
Also of relevance, data have been obtained
for leading neutron production in proton-heavy ion ($pA$) and
heavy ion-heavy ion collisions (for recent data see, e.g. 
Barish et al. 2002, Huang et al. 2002, Armstrong et al. 1999)
and for $e^+ p$ collisions by the ZEUS Collaboration at Hera
\cite{Chekanov2002,Levman2002a,Levman2002b}.

In the astrophysics literature a wide
range of values for the leading neutron multiplicity
has been employed from $\sim 0.1$ \citep{Contopoulos1995} to $\sim 0.5$
(Berezinskii \& Gazizov 1977)
with values close to one quarter
perhaps most generally being favored (e.g., Sikora (1989); Jones (1990); Drury (1994)).
But, given that this preference seems
a matter of lore, one must really look to the data.

Unfortunately, here again the picture is not entirely
clear. From fits to the ISR data \citep{Engler1975,Flauger1976}, obtained at 
CMS energies between around 16 and 63 GeV,
various 
researchers have determined values for the leading neutron multiplicity, $n_n$,
between 0.25 and 0.40. 
Though we have not re-analysed the data ourselves,
our considered opinion is that
a neutron multiplicity at the higher end of this range is actually the best motivated.
In support of this position, note that, for instance,
 though (from the ISR data)
\citet{Frichter1997}  obtain $n_n = 0.27$, they also find a value for the 
leading proton multiplicity, $n_p$, of 0.62 which, together with the constraint that 
$n_p + n_n = 1$ (where leading hyperon production, as supported by the data,
can be  neglected approximately), is consistent with a value for the leading neutron 
multiplicity at the higher end of the above range
(and, of course, the proton is generically easier to track than the neutron). 
Also note that from the same data, Forti {\em et~al.} (1990)
find $n_n = .40$. 
Further, from lower energy experiments conducted at the 2m hydrogen bubble
at the CERN proton synchrotron in the late 1970s, 
Blobel {\em et~al.} (1978)
determine $\sigma_n/(\sigma_p + \sigma_n)$ to be $\sim 0.38$ and
$0.35$ at CMS energies of 12 and 24 GeV respectively (where
$\sigma_p$ and $\sigma_n$ are, respectively, the total, inclusive
cross-sections for $pp \to ppX$ and $pp \to pnX$).
As we set out below, the
more sophisticated theoretical treatments also predict a leading
neutron multiplicity closer to one half than one quarter. In conclusion, therefore,
we take it that the value of $n_n$ is 0.40 at ISR energies.

\subsection{Leading Neutron Elasticity in Proton-Proton Interactions: Experiment}

Again, there is a complete lack of direct, leading particle inelasticty 
measurements at the EeV energy scales we are concerned with and we can only look to the ISR data. 
For the Feynman
x variable
\be
x_F \equiv p_\parallel/p_{\parallel}^{max} 
\simeq 2p_\parallel/\sqrt s,
\ee
which, at the energies of interest we can take to be indentical to the inelasticity,
\citet{Frichter1997} find from the ISR data average values for protons and neutrons
of $\langle x_{F \, p} \rangle = 0.44$ and $\langle x_{F \, n} \rangle = 0.26$.
The fact that the data indicate that leading neutrons be softer than leading protons
is consistent with the expectation from theory: see below. 

\subsection{Theory at Low Energies and the Interpretation of the ISR Data}
If Feynman scaling were exactly obeyed,
the inclusive distribution of final state particles would be given by:
\begin{eqnarray}
{d\sigma_{pp \to a} \over dp_\parallel \, d^2 p_\perp} (p_\parallel, p_\perp, \sqrt{s})
&\simeq  & 
f_a(p_\parallel, \sqrt{s}) ~G_a (p_\perp)  \nonumber \\
& \simeq  &
{ F_a(x_F)\over \sqrt{s}} \; G_a(p_\perp),
\label{eqn_scaling}
\end{eqnarray}
where 
the  functions $F_a(x_F)$ and 
\be
\label{eqn_perp}
G_a (p_\perp)
 \propto e^{-bp_\perp^2}
\ee
are independent of $\sqrt{s}$ \citep{Lipari2003}.
The ISR data are, in fact, well described by such scaling:
the cross-section
for leading neutron production was found to fall off
exponentially in $p_\perp$ for fixed Feynman 
$x_F$
  and could
be parameterized by
equation (\ref{eqn_scaling})
with $b$ parameter of $(4.8 \pm 0.3)$ (GeV$/c)^{-1}$, independent of
$s$ and $x_F$ in the range $0.2 \geq x_F \geq 0.7$ \citep{Engler1975}. 
Further, the invariant cross-section shows no dependence
on $s$ in the ISR data.

At the energies probed by this experiment, 
the dominant mechanism for leading neutron
production in pp is direct production
mediated by single pion
exchange
($\pi^+$).
The $\pi$ dominates the ISR-energy $p \to n$ transition amplitude
because of its low mass, its relative contribution
increasing for $|t|$ decreasing \citep{Chekanov2002}.
Other processes, which could otherwise be 
expected to contribute at the 20-30\% level---and which are 
required for the self-consistency of the meson exchange 
picture---in fact only contribute marginally (5\% level; 
Levman et al. 2002a).  
Such secondary processes include exchange of 
the $\rho$ and
$a_2$ mesons, off mass-shell effects 
and absorption.  
There can also be marginal contributions from 
two-meson exchange, and
different exchanges can interfere \citep{Levman2002a}.
Indirect production processes including diffractive
(or resonant) production of heavy baryons  also play a part
(i.e., $\Delta^+$ and
$\Delta^0$ resonances; these two
being excited baryon multiplets of $uud$ and
$udd$ quarks, respectively).

As set out above  it was demonstrated at the ISR
that leading neutrons are significantly softer than their proton counterparts.
This 
can be understood in the following way: the emergence of a leading neutron from a 
pp interaction requires that one of the incident protons has `swapped' a valence 
$u$ for a valence $d$. But the $u$ that has been cast out may itself be hadronized into
a leading meson which would be expected to carry a large fraction of the initial momentum.
This suppresses leading neutron production at $x_F \to 1$.

\subsection{Extrapolating Multiplicity and Elasticity to High Energies: 
Gribov-Regge Theory and Montecarlo Modeling}

Clearly if Feynman scaling always held exactly one could easily extrapolate
low energy results to  predict  the properties of hadronic interactions at an arbitrarily
high
energy.
Data obtained on pp interactions in general (conducted since the ISR experiments) at
 Sp$\overline{\rm p}$S,
Tevatron and RHIC, however,  have shown that Feynman scaling is violated with,
in particular, both the afore-mentioned, slow cross-section growth
and some evolution of secondary, charged particle multiplicity evident in the data
\citep{Lipari2003}.
This can be naturally interpretted as symptomatic of the fact that at increasingly
high `probe' momenta, the detailed QCD structure of the hadron is progressively revealed
(i.e., the description of the proton as a bag of three valence quarks breaks down).

Unfortunately, the higher center-of-mass energy experiments mentioned above
(which, admittedly, are
still at center-of-mass energies below our region of concern) 
have not been able to probe $pp \to $ leading $n$ (for the general reasons outline above).
Experimental data, then,
on multiplicity and elasticity
must be extrapolated over fully
 three orders of magnitude (required to reach the super-EeV scales that concern us)
using strong interaction models
and {\it this extrapolation may not  assume Feynman scaling}.
But here, again, we run into problems because the leading
particle production processes are soft and, therefore, 
associated with long distances. They cannot, then, be studied with perturbative QCD.
Instead we must rely on Regge theory and the instantiation
of concepts like the Pomeron in Monte Carlo codes (see below).

A general framework to effect the extrapolation of 
hadronic interactions from collider energies
to the super-EeV energy scales we are interested in has been realized over the last decade
\citep{Werner1993, Engel2001b, Ostapchenko2003} 
in a number of
different Montecarlo codes.
These are based on
Gribov's Reggeon Field Theory or the QCD eikonal approach \citep{Ostapchenko2003}.
In both these, a single 
hadronic interaction can be broken down into
a set of pomeron exchanges between participating partons 
(Gribov 1968; Engel 2001; Lipari 2003)
\footnote{Note that while
single-pomeron exchange generally predict results indistinguishable from
these models 
at energies up to several TeV in the center-of-mass, the single-pomeron
exchange amplitude eventually violates unitarity. Multiple pomeron
exchange, however, predicts a weaker energy growth above this energy scale and cures the 
unitarity problem: \cite{Engel2001}.}. 
As explained above, the increasing number of parton-parton interactions that open
up with increasing $\sqrt s$ explain the weak energy dependence
(and, therefore Feynman-scaling violation) seen in the cross-section, multiplicity, etc., data.
Further, an approach involving
muliple particle scattering, seems to be
the requirement if one is to arrive at a 
self-consistent description of Feynman scaling violations in secondary
particle spectra \citep{Ostapchenko2003}. This topic is beyond the scope of this work; 
we simply quote results obtained by other researchers below.


\subsubsection{Leading Neutron Multiplicity at High Energies}

In modeling of non-diffractive production, the ratio of leading proton to leading
neutron reflects the proton fragmentation scheme employed \citep{Huang2002}. Again,
if scaling were exactly obeyed this ratio would show no dependence on $\sqrt s$.
More complicated models predict a small energy dependence in this 
ratio\footnote{Paolo Lipari 2003, private communication.} (this is also the case for
diffractive production). Such energy dependence as does arise is due to
the increase in the number of sub-interactions between partons that occurs
for increasing $\sqrt s$ \citep{Lipari2003}.
In a simple diquark-quark proton fragmentation scheme, the incident proton can fragment
into either a $uu$ or a $ud$ diquark. 
The $uu$ must hadronize into a proton, but the
$ud$ can fragment into either a proton or a neutron.
Thus, this simple scheme predicts a proton-to-neutron ratio of 2:1 for leading 
baryons (so $n_n$ would be $\sim 0.33$). More sophisticated 
schemes---like three-quark fragmentation and gluon-junction interaction
models---predict, however, ratios closer to 1:1 \citep{Huang2002}.

\subsubsection{Leading Neutron Elasticity at High Energies}

In regard to the simulation of
UHE particle production with Monte Carlo codes, particular regard has been paid to
leading neutron production by Capdevielle, Attallah and co-workers
\citep{Capdevielle1992a,Capdevielle1992b}.
Employing a Monte Carlo generator developed for the simulation of
cosmic ray cascades on the basis of a simplified picture of the dual
parton model \citep{Capella1994}, these authors have investigated neutron 
production
up to the centre of mass energy regime
of $\sqrt s = 50$ TeV that we are interested in.
Their results incorporate direct production of leading neutrons and
diffractive
production through the $\Delta^+$ and $\Delta^0$. In addition
to the leading neutrons, they simulate production of
neutron secondaries with small rapidity 
in nucleon-anti-nucleon pairs and through
heavy baryon production and decay ($\Lambda$'s, $\Xi$'s and
$\Sigma$'s: \citep{Capdevielle1992b}). On the basis
of an extrapolation of UA5 data, Capdevielle and Attallah determine that 
the average energy carried away
by a leading neutron is 23-24\% that of the incoming energy.
This figure is very close to the neutron elasticity
obtained
from the ISR data, and, therefore
consistent with (at worst) a weak energy-dependence that we may disregard. 

\subsection{Conclusion: Final Assumptions on Leading Neutron Multiplicty and Inelasticity at EHE}

In summary, we have the following:
\begin{enumerate}
\item
Data indicate that the total (inelastic) cross-section
for pp interactions grows slowly with $\sqrt{s}$ so that scaling is violated. We
account for this effect in this paper.
\item
The highest energy experimental data available indicate, for pp interactions,
 a leading neutron multiplicity expectation of $n_n = 0.4$ and an averaged
leading neutron elasticity of $\langle x_n \rangle = 0.25$. 
\item
Modeling
indicates that $n_n$ and $\langle x_n \rangle$ are, at worst, only weakly dependent on
CMS energy. As our final position, then,
we adopt the above values for our calculations
and take them to hold independent of $s$.

\end{enumerate}

\section{$\chi^2$ Fitting Function}

For the sake of completeness, we set out the $\chi^2$ functions we have 
defined to provide a quantitative estimate of the goodness-of-fit of
our model of the totality of the data (i.e., EGRET, HESS and EHE cosmic ray measurements).
For particular instances 
described in the body of the paper
we define an overall $\chi^2$ with contributions, as appropriate,
from elements of
\be
\left\{ \chi^2_\mathrm{EGRET} \;, \;\chi^2_\mathrm{HESS} \;,\;  \chi^2_\mathrm{Neutron} \right\}\; .
\ee
The total $\chi^2$ is minimized by fitting to the spectral index and to the
natural logarithm of the normalization of the $\gamma$-ray  differential flux
which is assumed to be governed by a power law.
We work in logarithms because this means that, to a good approximation, the
fitted quantities are linear in the fitting parameters. We use the result that, for a linear model
$y(x) = a + bx$, in which the experimental limitations mean that both 
the observed quantities ($y_i's$) and
the parameter values {\it at} which they are observed
($x_i's$) are uncertain, a $\chi^2$ may be defined by
\be
\chi^2 = \sum_i \frac{(y_i - a - b x_i)^2}{\sigma^2_{y_i} + b^2 \sigma^2_{x_i}} \ .
\ee 
We find, then,
(we disregard the units of naturally dimensionful quantities and take all 
energies to be measured in 
eV, all fluxes in  cm$^{-2}$ s$^{-1}$, and all differential fluxes in 
cm$^{-2}$ s$^{-1}$ eV$^{-1}$):
\be
\chi^2_\mathrm{EGRET} = 
\sum_i \frac{\left(\log\left[\frac{d F_\mathrm{EGRET}^{obs}(E_i)}{dE}\right] - \log[Norm] + \gamma \log[E_i]\right)^2}
{\left( \Delta \left[\log\left[\frac{d F_\mathrm{EGRET}^{obs}(E_i)}{dE}\right]\right]\right)^2 + 
\gamma^2 \left(\Delta\left[\log[E_i]\right] \right)^2}
\ee
and
\be
\chi^2_\mathrm{Neutron} = \frac{(\log[F_n^{obs}] - \log[F_n^{pred}(\gamma, norm)])^2}
{(\Delta[\log[F_n^{obs}]])^2 + (1 - \gamma)^2 (\Delta[\log[E_\mathrm{Neutron}]])^2} \ .
\ee
The HESS $\chi^2$ has the same form as the EGRET $\chi^2$.
There are a number of points to note in relation to these equations. Firstly,
$Norm$ -- a number in cm$^{-2}$ s$^{-1}$ eV$^{-1}$ -- denotes the fitted,
\gam -ray, differential flux normalization. Secondly,
$F_n^{pred}$ in the last equation is as given by Equations (\ref{eqn_neut_flux_adjusted}) and (\ref{eqn_diff_flux_neut}).
Thirdly, strictly the $\chi^2$ analysis only holds for the case of Gaussian statistics in which 
case errors should be symmetric about the measured values 
(of the logarithms of the energies or fluxes). Our data do not always precisely satisfy this criterion. 
In the cases
where they do not, we take, then, the geometrical mean of the 
ratios (best value):(smallest value) and (largest value):(best value) to define the error.
Lastly, the sum in the EGRET $\chi^2$ is over the last three data points obtained from this instrument's GC
observation (this is to guarantee that we are in the scaling regime so that there be no curvature in the
photon differential flux).

\section{Derivation of the Maximum Energy}

In this appendix we present the derivation of equation (\ref{eqn_E_max_perp}).
From Eqs. (\ref{eqn_pc}) and (\ref{eqn_kap_rat}), we have for a perpendicular shock
\bea
\label{eqn_pc_perp}
\frac{1}{p_c}\frac{\textrm{d} p_c}{\textrm{d} t } 
&\simeq& \frac{ U_\mathrm{shock}^2}{ 4 \kappa_\perp}\nn \\
&=&\frac{ U_\mathrm{shock}^2}{ 4 \kappa_\parallel}
\left[1 + \left(\frac{\lambda_\parallel}{r_g} \right)^2 \right] \nn \\
&=& \frac{3}{4 \lambda_\parallel} \frac{U_\mathrm{shock}^2}{w}
\left[1 + \left(\frac{\lambda_\parallel}{r_g} \right)^2 \right] \nn \\
&\simeq& \frac{3}{4} \frac{U_\mathrm{shock}^2 \lambda_\parallel}{w \, r_c^2}
\eea
We again take the maximum value of $\Upsilon$  $\sim 0.3$, so that
\be
\label{eqn_Up_max}
\frac{U_\mathrm{shock}} {w}\frac{\lambda_\parallel}{r_c} \to 0.3,
\ee
so that substituting equation (\ref{eqn_Up_max}) into equation (\ref{eqn_pc_perp}) we find
\be
\frac{1}{p_c}\frac{\textrm{d} p_c}{\textrm{d} t }  \simeq \frac{1}{4} \frac{U_\mathrm{shock}}{r_c}.
\ee
Noting that the gyroradius is
\be
r_g = \frac{p_\perp}{e B}\;,
\ee
we have that $r_c \simeq p_c/eB$.  Therefore, 
\be
p_c\simeq{eB\over 4}\int U_{\mathrm shock}\,dt\;,
\ee
or to put this in a more observationally-motivated form,
\be
p_c\simeq{eB\over 4}\,R_{\mathrm shock}\;,
\ee
where $R_{\mathrm shock}$ is the characteristic shock radius.
Numerically, 
\be
p_c\simeq 5\times 10^{16}\;Z\left({B\over 20\;\mu G}\right)\left({R_{\mathrm shock}\over
10\;\mathrm{pc}}\right)\; \frac{\mathrm{eV}}{c}.
\label{eqn_p_max}
\ee





\end{document}